\documentclass[11pt]{article}
\pdfoutput=1

\usepackage{pstricks}
\usepackage{pst-plot}
\usepackage{pst-node}
\usepackage{amsmath}
\usepackage{pst-grad}
\usepackage{amssymb}
\usepackage{xspace}
\usepackage{pst-math}
\usepackage{pst-xkey}
\usepackage{pst-3dplot}
\usepackage[T1]{fontenc}
\usepackage{ae}
\usepackage[ansinew]{inputenc}
\usepackage{graphics}
\usepackage{color}
\usepackage{float}
\usepackage{graphicx}

\newrgbcolor{darkred}{0.8 0 0}
\newrgbcolor{darkerred}{0.7 0 0}
\newrgbcolor{darkestred}{0.5 0 0}
\newrgbcolor{darkgreen}{0 0.5 0.5}
\newrgbcolor{darkergreen}{0 0.6 0}
\newrgbcolor{darkestgreen}{0 0.5 0}
\newrgbcolor{darkblue}{0 0 0.6}
\newrgbcolor{darkestblue}{0 0 0.5}
\newrgbcolor{redgreen}{0.8 0.8 0}
\newrgbcolor{bluegreen}{0 0.4 0.7}
\newrgbcolor{bluewhite}{0 0.9 1}
\newrgbcolor{lightgreen}{0.6 1 0.6}
\newrgbcolor{lightred}{1 0.8 0.8}
\newrgbcolor{lightyred}{1 0.4 0.4}
\newrgbcolor{lightyblue}{0.5 0.5 1}
\newrgbcolor{lightblue}{0.8 1 1}
\newrgbcolor{lilas}{0.5 0.3 0.7}
\newrgbcolor{lilaswhite}{1 0.8 1}
\newrgbcolor{redy}{1 0.6 0.6}
\newrgbcolor{bluish}{0.6 0.6 1}
\newrgbcolor{sand}{0.9 0.9 0.7}
\newrgbcolor{brown}{0.4 0.4 0}
\newrgbcolor{blue1}{0 0.35 0.65}
\newrgbcolor{blue2}{0 0.4 0.6}
\newrgbcolor{blue3}{0 0.45 0.55}
\newrgbcolor{blue4}{0 0.5 0.5}
\newrgbcolor{blue5}{0 0.55 0.45}
\newrgbcolor{blue6}{0 0.6 0.4}
\newrgbcolor{blue7}{0 0.65 0.35}

\begin{document}

\title{Dynamics in two networks based on stocks of the US stock market}

\author{Leonidas Sandoval Junior \\ \\ Insper, Instituto de Ensino e Pesquisa \\ Rua Quatá, 300, São Paulo, SP, Brazil, CEP 04546-042 \\ E-mail: leonidassj@insper.edu.br \\ Telephone number: (55) (11) 45042300}

\maketitle

\begin{abstract}
We follow the main stocks belonging to the New York Stock Exchange and to Nasdaq from 2003 to 2012, through years of normality and of crisis, and study the dynamics of networks built on two measures expressing relations between those stocks: correlation, which is symmetric and measures how similar two stocks behave, and Transfer Entropy, which is non-symmetric and measures the influence of the time series of one stock onto another in terms of the information that the time series of one stock transmits to the time series of another stock. The two measures are used in the creation of two networks that evolve in time, revealing how the relations between stocks and industrial sectors changed in times of crisis. The two networks are also used in conjunction with a dynamic model of the spreading of volatility in order to detect which are the stocks that are most likely to spread crises, according to the model. This information may be used in the building of policies aiming to reduce the effect of financial crises.
\end{abstract}


\vskip 0.3 cm


\vskip 0.3 cm

\noindent {\bf Some figures were removed from this version due to the large size (in bits) of them. For a full version, please download from

\noindent https://insper.academia.edu/LeonidasSandoval/Papers?s=nav\#add .}

\section{Introduction}

The issue of the spreading of crises among financial markets has been a topic of intensive study, mainly after the crisis of 2008 and the subsequent crises. The mapping of the network of banks and other financial institutions is now considered essential to the understanding of how defaults can propagate from one institution to another, and understanding the network of financial institutions has been placed among the main challenges of the present (Haldane, 2009). This article contributes to a better understanding of a network based on stocks negotiated in the two main stock exchanges of the USA, that are also among the largest stock markets in the world: the New York Stock Exchange and the Nasdaq. In order to do so, we use two measures of how one stock relates to another: the Pearson correlation and Transfer Entropy.

The Pearson correlation, developed by Karl Pearson (1857–1936), measures how similar are the time series of two variables (stocks, in our case). Transfer Entropy (TE), developed by Thomas Schreiber (2000), is a measure of the amount of information that the time series of one variable has on the time series of another variable that was not already in the time series of the latter. The first measure, correlation, is symmetric and based on linear relations, although there are correlation measures, like the Spearman rank correlation and the Kendall tau rank correlation, that measure nonlinear correlations. The second measure, Transfer Entropy, is dynamic, non-symmetric, and is related to Granger causality, although it is not model-dependent, and is capable of detecting nonlinear relations between variables. Both measures are applied to 464 stocks negotiated in the New York Stock Exchange and/or Nasdaq, from 2003 to 2012, in order to build networks of stocks. The evolution of these networks are then studied through time, and the dynamics of the relationships between stocks and between sectors are studied both from the point of view of similarity (using correlations) and of causality (using TE). The networks are also used, in conjunction with a model for the spreading of volatility, in order to detect, according to the model, which are the stocks more likely to spread crises.

There is an extensive literature on the propagation of shocks in networks of financial institutions, and describing all the published works in this subject is beyond the scope of this article. Most of the works in this field can be divided into theoretical and empirical ones, most of them considering networks of banks where the connections are built on the borrowing and lending between them. In most theoretical works (Kirman, 1997, Allen and Gale, 2000; Watts, 2002; Vivier-Lirimont, 2004; Leitner, 2005; Nier, Yang, Yorulmazer and Alentorn, 2007; Castiglionesi and Navarro, 2007; Cossin and Schellhorn, 2007; Lorenz, Battiston and Schweitzer, 2009; Schweitzer, Fagiolo, Sornette, Vega-Redondo, and White, 2009; Allen and Babus, 2009; Gai and Kapadia, 2010; Georg, 2010; Canedo and Martínez-Jaramillo, 2010; Gai, Haldane and Kapadia, 2011; Tabak, Takami, Rocha, and Cajueiro, 2011; Battiston, Gatti, Gallegati, Greenwald, and Stiglitz, 2012a; Battiston, Gatti, Gallegati, Greenwald, and Stiglitz, 2012b; Amini, Cont, and Minca, 2012; Elliott, Golub and Jackson, 2013; Acemoglu, Osdaglar and Tahbaz-Salehi, 2013), networks are built according to different topologies (random, small world, or scale-free), and the propagation of defaults is studied on them. The conclusions are that small world or scale-free networks are, in general, more robust to cascades (the propagation of shocks) than random networks, but they are also more prone to propagations of crises if the most central nodes (usually, the ones with more connections) are not themselves backed by sufficient funds.

Most empirical works (Boss, Elsinger, Summer, and Thurner, 2004; Müller, 2006; Soramäki, Bech, Arnold, Glass, and Beyeler, 2007; Hattori and Suda, 2008; Iori, Masi, Precup, Gabbi, and Caldarelli, 2008; Markose, Giansante, Gatkowski, and Shaghaghi, 2010; Kubelec and Sá, 2010; Minoiu and Reyes, 2011; Lee, Yang,  Kim, Lee, Goh, and Kim, 2011; Upper, 2011; Battiston, Puliga, Kaushik, Tasca, and Caldarelli, 2012; Martínez-Jaramillo, Alexandrova-Kabadjova,Bravo-Benítez, and Solórzano-Margain, 2012; Hale, 2012; Kaushik and Battiston, 2012; Chinazzi, Fagiolo, Reyes, and Schiavo, 2013; Memmel and Sachs, 2013) are also based on the structure derived from the borrowing and lending between banks, and they show that those networks exhibit a core-periphery structure, with few banks occupying central, more connected positions, and others populating a less connected neighborhood. Those articles showed that this structure may also lead to cascades if the core banks are not sufficiently resistant, and that the network structures changed considerably after the crisis of 2008, with a reduction on the number of connected banks and a more robust topology against the propagation of shocks.

Section 2 of this article explains the data used. Section 3 uses the correlations between stocks in order to study the similarity of behavior in our data set. Section 4 explains Transfer Entropy and uses it in order to study the exchange of information between stocks. Section 5 builds two networks, one based on correlations, and another built on TE, and makes a study of the centrality of stocks according to complex networks theory. Section 6 studies the dynamics of both networks through times of crises, Section 7 makes simulations with shocks originating both in one particular stock and shocks exogenous to the system which affect all stocks or some sector of the economy, and Section 8 presents some conclusions.

\section{Data}

We work with the stocks belonging to the index S\&P 500 of the New York Stock Exchange (500 stocks in the index) and the Nasdaq 100 index of Nasdaq stock exchange (100 stocks in the index). Only the stocks with a certain liquidity were considered, what means they were negotiated in nearly all the days (more than 80\%) the stock exchanges operated, what eliminates a small number of stocks. There is a large intersection of stocks that are negotiated in both stock exchanges, and we eliminated any duplicate data from our sample, ending with 464 stocks. The stocks were ordered according to a sector classification used by Bloomberg, where the data was taken from. The sectors are Basic Materials (26 stocks), Energy (37 stocks), Industrial (76 stocks), Consumer, Cyclical (62 stocks), Diversified (1 stock), Financial (71 stocks), Communications (34 stocks), Technology (54 stocks), Utilities (29 stocks), and Consumer, Non-Cyclical (94 stocks).

The sectors and industries are organized in Table 1. The order of sectors is such that the most correlated sectors are close together. The company name, tickers, sector, industry, and sub-industry of each of the stocks in the data are detailed in Appendix A. Since there is just one company in the Diversified sector, a holding company that invests in a variety of sectors, we are placing it together wit the Financial sector when producing graphs or calculating aggregate data, for visual purposes.

\small

\[ \begin{array}{l|l} \text{\bf Sector} & \text{\bf Industries} \\ \hline \text{\bf Basic Materials} & \text{Chemicals, Forest Products and Paper, Iron/Steel, Mining, Quarrying.} \\ \text{\bf Energy} & \text{Coal, Oil \& Gas, Oil \& Gas Services, Pipelines.} \\ \text{\bf Industrial} & \text{Aerospace/Defense, Building Materials, Electrical Components \& Equipment,} \\ & \text{Electronics, Engineering \& Construction, Hand/Machine Tools,} \\ & \text{Machinery-Construction \& Mining, Machinery-Diversified,} \\ & \text{Metal Fabricate/Hardware, Miscellaneous Manufacturing, Packaging \& Containers,} \\ & \text{Transportation.} \\ \text{\bf Consumer, Cyclical} & \text{Airlines, Apparel, Automanufacturing, Autoparts \& Equipment,} \\ & \text{Distribution/Wholesale, Entertainement, Home Builders, Home Furnishing,} \\ & \text{Houseware, Leisure Time, Lodging, Retail, Textiles, Toys/Games/Hobbies.} \\ \text{\bf Diversified} & \text{Holding Companies-Diversified.} \\ \text{\bf Financial} & \text{Banks, Diversified Financial Services, Insurance, REITS, Savings \& Loans.} \\ \text{\bf Communications} & \text{Advertising, Internet, Media, Telecommunications.} \\ \text{\bf Technology} & \text{Computers, Electronic Components \& Equipment, Electronics,} \\ & \text{Office/Business Equipment, Semiconductors, Software.} \\ \text{\bf Utilities} & \text{Electric, Gas.} \\ \text{\bf Consumer, Non-Cyclical} & \text{Agriculture, Beverages, Biotechnology, Commercial Services,} \\ & \text{Cosmetics/Personal Care, Food, Healthcare Products, Healthcare Services,} \\ & \text{Household Products/Wares, Pharmaceuticals.}

\end{array} \]

\normalsize

\noindent {\bf Table 1.} Sectors and industries as classified by Bloomberg.

\vskip 0.3 cm

The daily closing prices of each stock are used in order to calculate log-returns, given by
\begin{equation}
\label{logrets}
R_t=\ln (P_t)-\ln (P_{t-1})\ ,
\end{equation}
where $P_t$ is the closing price of the stock at day $t$ and $P_{t-1}$ is the closing price of the same stock at day $t-1$. We worked with the log-returns in order to avoid issues due to the nonstationarity of the time series of the closing prices. Working with end of trading day returns, we are not studying the high frequency trading of the market, that drives prices during the day, but the slower dynamics of prices along longer periods. We will do some work with intraday data in the near future.

\section{Correlations}

Our first analysis of the data is based on the familiar correlation structure between the stocks in our sample. We use the Pearson correlation, which is given by
\begin{equation}
\label{Pearson}
C_{ij}=\frac{\sum_{k=1}^n\left( x_{ik}-\bar{x_i})(x_{jk}-\bar{x_j}\right)}{\sqrt{\sum_{k=1}^n\left( x_{ik}-\bar{x_i}\right) ^2}\sqrt{\sum_{i=1}^n\left( x_{jk}-\bar{x_j}\right) ^2}}\ ,
\end{equation}
where $x_{ik}$ is element $k$ of the time series of variable $x_i$ and $x_{jk}$ is element $k$ of the time series of variable $x_j$, and $\bar{x_i}$ and $\bar{x_j}$ are the averages of both time series, respectively.

The Pearson correlation is used in order to calculate the linear correlation between variables. Other types of correlation measures, like the Spearman rank correlation and the Kendall tau rank correlation, are used in order to calculate nonlinear relations between variables. Here, we apply the usual Pearson correlation because it has been shown (Sandoval, 2013) that the results using this correlation measure are very similar to the Spearman rank correlation for the financial data we are using and is much faster to compute.

The structure of the resulting correlation matrix may be visualized in Figure 1 (left, where on the right the same picture is plotted with the sectors highlighted), where we plot a heat map of the elements of the correlation matrix, with lighter colors denoting higher correlations and darker colors denoting lower correlations. The figure displays the correlations in such a way that the leftmost and lowest corner corresponds to the correlation between element 1 with itself. The number of each stock grows from left to right and from the bottom to the top. The same configuration will be used in all other representations of matrices in this article. As expected, the diagonal elements are the brightest ones, with correlation 1 between all stocks and themselves. It is also possible to identify some clusters, related with sector and industry of stocks.

One can notice some black or dark lines, corresponding to stocks that do not correlate well with other stocks in the sample. For Basic Materials, there are clusters and subsclusters, corresponding to industries; Energy is a large, compact cluster; Industrial has many subclusters based on industries, but all of them very sparse; inside Consumer, Cyclical, there are some sparse clusters and a compact one in Distribution/Wholesale; Financial has a sparse subcluster for Banks and a compact subcluster for REITS; for Communications, there is an inner cluster corresponding to the industry Media; Technology has a subcluster of the Semiconductors industry; Utilities form a compact cluster; and for Consumer, Non-Cyclical, there is a sparse subcluster for the industry Health Care Services and two very sparse ones for Healthcare Products and for Pharmaceutilcals.

Values in Figure 1 go from -0.0429 (slightly anticorrelated) to 1 (totally correlated), with a maximum of 0.9064 if we exclude the self-correlations (which are always equal to 1). We may compare these values with the ones that may be obtained by considering all time series of data, but randomly shuffling each time series independently, so that any true correlation between the time series is broken, but the probability distribution of each one is maintained. By computing 1,000 simulations randomizing the time series and then calculating the correlation matrix for each simulation leads to a minimum correlation $-0.10\pm0.01$ (average $\pm $ standard deviation) and a maximum correlation $0.10\pm0.02$ (excluding self-correlations). Figure 2 shows a histogram of the correlation matrix values and a histogram obtained from the simulations with randomized data, both with the autocorrelations removed. It is clearly visible that, except for a small quantity of classes, the correlation matrix for real data presents very distinct results from the correlations obtained with randomized data. So, the correlation between stocks is well above the correlation predicted for uncorrelated data.

\section{Transfer Entropy}

Although useful for determining which stocks behave similarly to others, the correlations between them cannot establish a relation of causality or of influence, since the action of a stock on another is not necessarily symmetric. A measure that has been used in a variety of fields, and which is both dynamic and non-symmetric, is {\sl Transfer Entropy}, developed by Schreiber (2000), which is based on the concept of {\sl Shannon Entropy}, first developed in the theory of information by Shannon (1948). Transfer entropy has been used in the study of cellular automata in Computer Science, in the study of the neural cortex of the brain, in the study of social networks, in Statistics, and also in the analysis of financial markets, as in the works of Marschinski and Kantz (2002), Baek, Jung, and Moon (2005), Kwon and Yang (2008a), Kwon and Yang (2008b), Jizba, Kleinert, and Shefaat (2012), Peter, Dimpfl, and Huergo (2012), Dimpfl and Peter (2012), Kim, An, Kwon, and Yoon (2013), Li, Liang, Zhu, Sun, and Wu (2013), Dimpfl and Peter (2014), and Sandoval (2014). In this section, we shall describe the concept of Transfer Entropy (TE), using it to analyze the data concerning the 464 stocks in our sample and their lagged counterparts.

When one deals with variables that interact with one another, then the time series of one variable $Y$ may influence the time series of another variable $X$ in a future time. We may assume that the time series of $X$ is a Markov process of degree $k$, what means that a state $i_{n+1}$ of $X$ depends on the $k$ previous states of the same variable. This may be made more mathematically rigorous by defining that the time series of $X$ is a Markov state of degree $k$ if
\begin{equation}
\label{Markov}
p\left( i_{n+1}|i_n,i_{n-1},\cdots ,i_0\right) =p\left( i_{n+1}|i_n,i_{n-1},\cdots ,i_{n-k+1}\right) \ ,
\end{equation}
where $p(A|B)$ is the conditional probability of $A$ given $B$, defined as
\begin{equation}
\label{conditional}
p(A|B)=\frac{p(A,B)}{p(B)}\ .
\end{equation}
What expression (\ref{Markov}) means is that the conditional probability of state $i_{n+1}$ of variable $X$ on all its previous states is the same as the conditional probability of $i_{n+1}$ on its $k$ previous states, meaning that it does not depend on states previous to the $k$th previous states of the same variable.

One may assume that state $i_{n+1}$ of variable $X$ also depends on the $\ell $ previous states of variable $Y$. The concept is represented in Figure 3, where the time series of a variable $X$, with states $i_n$, and the time series of a variable $Y$, with states $j_n$, are identified.

We may now define the concept of TE from a time series $Y$ to a times series $X$ as the average information contained in the source $Y$ about the next state of the destination $X$ that was not already contained in the destination's past. We assume that element $i_{n+1}$ of the time series of variable $X$ is influenced by the $k$ previous states of the same variable and by the $\ell $ previous states of variable $Y$. The values of $k$ and $\ell $ may vary, according to the data that is being used, and to the way one wishes to analyze the transfer of entropy of one variable to the other.

Transfer Entropy from variable $Y$ to variable $X$ is defined as
\begin{eqnarray}
\label{transferentropy}
TE_{Y\rightarrow X}(k,\ell ) & = & \sum_{i_{n+1},i_n^{(k)},j_n^{(\ell )}}p\left( i_{n+1},i_n^{(k)},j_n^{(\ell )}\right) \log_2p\left( i_{n+1}|i_n^{(k)},j_n^{(\ell )}\right) \nonumber \\ & & -\sum_{i_{n+1},i_n^{(k)},j_n^{(\ell )}}p\left( i_{n+1},i_n^{(k)},j_n^{(\ell )}\right) \log_2p\left( i_{n+1}|i_n^{(k)}\right) \nonumber \\
& = & \sum_{i_{n+1},i_n^{(k)},j_n^{(\ell )}}p\left( i_{n+1},i_n^{(k)},j_n^{(\ell )}\right) \log_2\frac{p\left( i_{n+1}|i_n^{(k)},j_n^{(\ell )}\right) }{p\left( i_{n+1}|i_n^{(k)}\right) }\ ,
\end{eqnarray}
where $i_n$ is element $n$ of the time series of variable $X$ and $j_n$ is element $n$ of the time series of variable $Y$, $p(A,B)$ is the joint probability of $A$ and $B$, and
\begin{equation}
p\left( i_{n+1},i_n^{(k)},j_n^{(\ell )}\right) =p\left( i_{n+1},i_n,\cdots ,i_{n-k+1},j_n,\cdots ,j_{n-\ell +1}\right) \end{equation}
is the joint probability distribution of state $i_{n+1}$, of state $i_n$ and its $k$ predecessors, and the $\ell $ predecessors of state $j_n$, as in Figure 3.

This definition of TE assumes that events on a certain day may be influenced by events of $k$ and $\ell $ previous days. Since most empirical data on financial markets suggest that log-returns of the prices of stocks have low memory (what is not true for volatility), we shall assume that we have a Markov state of degree 1 with respect to variables $X$ and $Y$, what means that we will consider that only the previous day in the time series of $X$ and $Y$ contain some information on the time series of $X$ at some target day. By doing so, formula (\ref{transferentropy}) for the TE of $Y$ to $X$ becomes simpler:
\begin{equation}
\label{TE}
TE_{Y\rightarrow X}=\sum_{i_{n+1},i_n,j_n}p\left( i_{n+1},i_n,j_n\right) \log_2\frac{p\left( i_{n+1}|i_n,j_n\right) }{p\left( i_{n+1}|i_n\right) }=\sum_{i_{n+1},i_n,j_n}p\left( i_{n+1},i_n,j_n\right) \log_2\frac{p\left( i_{n+1},i_n,j_n\right) p\left( i_n\right) }{p\left( i_{n+1},i_n\right) p\left( i_n,j_n\right) }\ ,
\end{equation}
where we took $k=\ell =1$, meaning we are using lagged time series of one day, only.

Applying (\ref{TE}) to our sample of data, one obtains a TE matrix, depicted in Figure 4, on the left, and with the sectors highlighted, on the right. The figure represents the TE going from elements of the vertical axis to elements of the horizontal axis, from lines to columns. The first feature that may be observed is that, although TE is not symmetric, the TE matrix shown in Figure 4 is remarkably symmetric, although not completely so. The values of TE go from 0 (darker colors) to 0.4067 (brighter colors).

Since brighter colors imply larger TE and darker colors imply lower TE, one can notice regions of low TE being received by stocks of the Utilities sector (horizontal lines from 341 to 370), and also from this same sector to all others (vertical lines from 341 to 370). There is also little TE from stocks inside the Utilities sector. Comparing with Figure 1, we may see that the Utilities sector is very correlated within itself, but little correlated with other sectors. We also see low TE from the Basic Materials sector, particularly from the Chemicals industry, to all others, and also from the Oil Companies industry in the Energy sector, from the Aerospace/Defense industry in the Industrial sector, part of the Retail industry of the Consumer, Cyclical sector, and from many industries of the Consumer, Non-Cyclical sector, with similar results for the stocks that receive little TE from the others.

Another feature are the larger values of TE between stocks of the Financial sector with themselves and with stocks of the other sectors. This indicates an exchange of information going both from the time series of stocks belonging to the Financial sector to the other stocks and from all stocks to the ones of the Financial sector. There is slightly less TE flowing from REITS to other REITS, what is typical of time series that behave as a block (very highly correlated, as can be seen from Figure 1), and thus do not exchange much information among themselves. The same explanation may be given to the lack of TE between stocks of the Utilities sector and themselves. There is also a good amount of TE from stocks belonging to the Communications and Technology sectors to other sectors and to themselves, and isolated lines of TE from stocks of the Basic Materials sector (Mining industry) and of the Consumer, Cyclical sector (Home Builders industry).

The size of the bins used in the calculations of the probabilities in (\ref{TE}) changes the resulting Transfer Entropy (TE) values. In Figure 4, we use a binning size 0.02, what leads to a much larger number of bins and to a much longer calculation time, but also gives a better resolution, given the amount of data used (see Sandoval, 2014b, for a better discussion of the information of the binning size on TE).

Although useful, the picture obtained from the TE matrix is not very informative, as there is little TE from one stock to another in the same day if we use daily returns. One different approach, applied to stocks of the financial sector in many stock exchanges around the world (Sandoval, 2014b) and to stock market indices (Sandoval and Kenett, 2014), is to consider the lagged time series of all stocks, together with the original time series, in order to build a larger set of data, twice the size of the original one, where original and lagged stocks are considered as different variables. This approach was first developed by Sandoval (2014a) in order to study stock markets around the world, which do not operate at the same times, using correlation. When applied to TE, one obtains very interesting results, shown in Figure 5, which displays the TE from all stocks to all stocks, original and lagged, with original stocks represented first, and lagged stocks represented next. The resulting expanded TE matrix is clearly not symmetric, and it may be divided into four sectors. The first sector (Sector 11), the lower left corner of the TE matrix, represents the TE from original to original stocks, and is the same as the TE represented in Figure 4. Now the second sector (Sector 21), top left corner of the TE matrix, represents the TE from lagged variables to the original ones. This sector presents some interesting structure, with brighter colors indicating large values of TE from the time series of stocks on the day before to the time series of stocks of the next day. There is a bright diagonal line inside this sector, which is the TE from one stock to itself on the next day, what is to be expected from the definition used (\ref{TE}), and other bright regions representing the exchange of information between time series of consecutive days. The third sector (Sector 12), lower right corner, represents the TE from original to lagged indices, and it is mostly noise, what is to be expected, since the transfer of information from the future to the past is not physically possible. The last sector (Sector 22), top right corner, shows the TE from lagged to lagged variables, and is mostly similar to Sector 11.

Figure 6 offers a closer look at Sector 21, representing the TE from lagged variables to original ones. Here also, one may see that, although Transfer Entropy is an asymmetric measure, the graph is particularly symmetric. Now, one must have in mind that this is the Transfer Entropy from lagged data to original (unlagged) variables, so this is a part of the very asymmetric expanded TE matrix in Figure 5 which is rather symmetric itself.  The values go from 0.0440 to 2.8029, about seven times larger than the maximum TE for sector 11.

There is a remarkable resemblance between the TE from lagged to original variables and the correlation matrix obtained before (Figure 1). This may be the effect of TE from one day to the other leading to the uniformity of behavior of stocks on the next day. Again, one may see strong values of TE from stocks of the Mining and Quarrying industries of the Basic Materials sector to themselves, from stocks of the Energy sector to themselves, from stocks of the Banks and REITS industries of the Financial sector to themselves, and from stocks of the Semiconductors industry of the Technology sector to themselves.

The similarity between the TE matrix for Sector 21 and the correlation matrix is striking. Figure 7 shows the scatter plot between both measures, and one can notice a nonlinear relation between them. The dots forming a vertical line at the right of the figure are due to the autocorrelations and to the TE from lagged variables to their original counterparts. The Pearson correlation coefficient of both measures is 0.6030, the Spearman rank correlation is 0.5690, and the Kendall tau rank correlation is 0.4037. These correlations are quite similar when the main diagonals of both matrices are removed. Besides the terms of autocorrelation and transfer entropy from one variable to itself on the next day, there are two groups that may be identified. The first one is for correlations close to zero, which are correlated with TE values that are either small or medium; the second one seems to be a relation between correlation and the square of TE.

This similarity between both measures indicate that the information sent from one stock to another stock on the next day and vice-versa is followed by a similar behavior of both stocks on this next day. So, the exchange of information between stocks seems to lead to a larger correlation between them on the next day.

Analyzing now the internal structures of the TE matrices for sectors 11 and 21, and comparing them with TE matrices obtained from randomized data, we may obtain an interval of validity of our results if compared with results of unrelated data. We made 10 simulations by flushing the data in each time series randomly, so as to destroy any relations between time series but maintain each probability distribution. We used just ten simulations because of the long computational times necessary to perform them, and also because there is very little difference between one simulation and another. Comparing the probability distributions of real data and of the randomized data, we obtain the probability distributions in Figure 8, where the left graph shows the results for sector 11 and the right graph displays results for sector 21.

Most of the values for sector 11 of the TE expanded matrix fall into the possible values for randomized data, so that most results may be seen as probably resulting from pure statistical noise, but results for sector 21 clearly detach themselves from results obtained from randomized data, indicating that there is a substantial amount of information that could not possibly be generated by statistical noise.

\section{Network structure}

Now we will build a network structure for the stocks based on their correlation matrix. This will be done by representing each stock in our sample as a node and correlations between them as edges. One way to do that is to represent the nodes in an abstract space where the distances between them are associated with their correlations: stocks that are more correlated appear as nodes that are closer together, and stocks that are less correlated appear as nodes that are farther apart from each other.

There are many ways to define a distance measure based on a correlation matrix, but the most used one in applications to financial markets is given by Mantegna (1999):
\begin{equation}
\label{distance}
d_{ij}=\sqrt{2\left( 1-C_{ij}\right) }\ ,
\end{equation}
where $C_{ij}$ is the correlation between nodes $i$ and $j$. As correlations between stocks vary from $-1$ (anticorrelated) to $1$ (completely correlated), the distances between them vary from $0$ (totally correlated) to $2$ (completely anticorrelated). Totally uncorrelated stocks would have distance $1$ between them.

Based on the distance measures, $m$-dimensional coordinates are assigned to each stock using an algorithm called Classical Multidimensional Scaling (Borg and Groener, 2005), which is based on minimizing the stress function
\begin{equation}
\label{stress}
S=\left[ \frac{\displaystyle{\sum_{i=1}^n\sum_{j>i}^n\left( \delta_{ij}-\bar d_{ij}\right) ^2}}{\displaystyle{\sum_{i=1}^n\sum_{j>i}^nd_{ij}^2}}\right] ^{1/2}\ \ ,\ \ \bar d_{ij}=\left[ \sum_{a=1}^m\left( x_{ia}-x_{ja}\right) ^2\right] ^{1/2}\ .
\end{equation}
where $\delta_{ij}$ is 1 for $i=j$ and zero otherwise, $n$ is the number of rows of the correlation matrix, and $\bar d_{ij}$ is an m-dimensional Euclidean distance (which may be another type of distance for other types of multidimensional scaling). The outputs of this optimization problem are the coordinates $x_{ia}$ of each of the nodes, where $i=1,\cdots, n$ is the number of nodes and $a=1,\cdots ,m$ is the number of dimensions in an $m$-dimensional space. The true distances are only perfectly representable in $m$ dimensions, but it is possible for a network to be well represented in smaller dimensions. In the case of this article we shall consider $m=2$ for a 2-dimensional visualization of the network, being the choice a compromise between fidelity to the original distances and the easiness of representing the networks in a two dimensional medium.

Figure 9 shows the stocks represented as nodes, with legends identifying different sectors. There is a crowded center where the majority of nodes of Communications and Technology stocks concentrate. Also occupying part of the center and its neighborhood are the stocks of Consumer, Cyclical, Consumer, Non-Cyclical, Financial, and Industrial sectors. The stocks belonging to Basic Materials are spread around the Industrial and Energy sectors; the Energy and Utilities sectors are set clearly apart from the other sectors, and the one stock belonging to the Diversified sector is also apart from the other stocks.

Figure 9 shows only the nodes of the network built using correlation, since drawing connections among all nodes would make the figure very confusing, since each node is connected to all others. There are filtering methods that drastically reduce the number of connections, and one of them, asset graphs, will be discussed in Subsection 5.4.

For Transfer Entropy, we may again try to produce a map of the nodes according to distances between stocks. The problem now is that distance is a symmetric measure, and Transfer Entropy is not. Another problem is that TE is not normalized. We may correct the latter problem by defining a normalized version of TE by dividing each column by the value of the TE from the lagged variable to itself. So, the TE from one lagged variable to itself is, at maximum, 1. One must be aware that this is not the measure usually called {\sl Normalized Transfer Entropy} in the literature, which is calculated in a very different way, and is not used here.

By using the definition given by (\ref{distance}), we may calculate a matrix for which the main diagonal is zero, but this matrix is still not symmetric, as a distance matrix must be. We chose to symmetrize the matrix by setting $d_{ij}=d_{ji}$ if $d_{ij}>d_{ji}$ and $d_{ji}=d_{ij}$, otherwise, what means that we always consider the smallest between the two values $d_{ij}$ and $d_{ji}$ to be the distance between $i$ and $j$. The resulting distance matrix is then used, applying (\ref{stress}), in order to calculate a set of coordinates for each stock as a node in a space where distances are similar to the ones given by the symmetrized distance matrix.

Figure 10 shows the two dimensional figure that results from this procedure, where stocks are represented as nodes colored according to sector, which is very similar to the graph obtained from correlations. The distances between nodes in the graph represent the best approximation to the distances calculated using the distance matrix based on Transfer Entropy.

Other choices for normalization or symmetrizing would lead to different graphs, but only slightly different. From Figure 10, one may see a detachment of the stocks belonging to the Energy, Financial, and Utilities industries. The other stocks seem to concentrate in a large cluster, although they maintain some coherence inside that cluster.

\subsection{Node Strength}

The correlation and Transfer Entropy networks produced by the correlation matrix of stocks and the TE between lagged and original stocks are {\sl weighted networks}, what means that the edges between nodes (stocks) have values attached to them, which are the correlations or the TE relations between stocks. For such networks, the main measure of the centrality of a node (in many ways, of its importance in the network) is {\sl Node Strength} ($NS$) (Newman, 2010), which for an undirected network such as the one obtained through correlation is the sum of all weights of the edges of a node (stock) with other nodes (stocks),
\begin{equation}
\label{NS}
{NS}_i=\sum_{j=1}^NC_{ij}\ ,
\end{equation}
where $C_{ij}$ is element $(i,j)$ of the correlation matrix, and $N$ is the number of stocks in our sample (464 stocks).

Since Transfer Entropy is asymmetric, there is a difference between the Transfer Entropy from one stock to all others and the Transfer Entropy from all stocks to one stock, so that the network formed using TE is a directed one. For directed networks, Node Strength assumes two guises: In Node Strength (${NS}_{in}$), which is the sum of the weights of all edges that go from all nodes to a particular node, and Out Node Strength (${NS}_{out}$), which is the sum of the weights of a node to all other nodes (Newman, 2010),
\begin{equation}
{NS}_{in}^i=\sum_{j=1}^N{TE}_{ij}\ \ ,\ \ {NS}_{out}^j=\sum_{i=1}^N{TE}_{ij}\ .
\end{equation}

Table 2 shows the values of the Node Strength for the network based on correlations for the ten stocks with highest centrality values. The stocks of Financial and Chemical companies occupy the first places, what means that those stocks have a behavior more similar to other stocks. Stocks of the Industrial sector and one stock of the Consumer, Cyclical sector follow.

\scriptsize
\[ \begin{array}{c|l|l|l} \hline \text{\bf Node Strength} & \text{\bf Company} & \text{\bf Sector} & \text{\bf Industry} \\ \hline 242.52 & \text{Franklin Resources Inc} & \text{Financial} & \text{Diversified Financial Services} \\ 241.31 & \text{T. Rowe Price Group Inc} & \text{Financial} & \text{Diversified Financial Services} \\ 239.54 & \text{DoubleDragon Properties Corp} & \text{Basic Materials} & \text{Chemicals} \\ 239.14 & \text{PPG Industries Inc} & \text{Basic Materials} & \text{Chemicals} \\ 235.54 & \text{Emerson Electric Co} & \text{Industrial} & \text{Electrical Components  \&  Equipment} \\ 234.84 & \text{Sigma-Aldrich Corp} & \text{Basic Materials} & \text{Chemicals} \\ 234.15 & \text{PACCAR Inc} & \text{Consumer, Cyclical} & \text{Auto Manufacturers} \\ 232.87 & \text{Loews Corp} & \text{Financial} & \text{Insurance} \\ 232.82 & \text{Dover Corp} & \text{Industrial} & \text{Miscellaneous Manufacturing} \\ 232.78 & \text{United Technologies Corp} & \text{Industrial} & \text{Aerospace / Defense} \\ \hline \end{array} \]

\normalsize

\vskip 0.2 cm

\noindent {\bf Table 2.} Classification of stocks with highest Node Strength, their sector and industry classifications. Only the ten stocks with highest centrality values are shown.

\vskip 0.3 cm

Table 3 represents the centrality results for the network based on TE, showing the top 10 stocks according to in or out Node Strength. All top 10 companies whose stocks receive the most information from all other stocks belong to the Financial sector, and mainly to the Insurance industry, what is understandable, since the prices of stocks belonging to insurance companies are highly influenced by the prices of companies they ensure. Now the companies whose stocks send the most information to all other stocks are more diverse, with two stocks of companies that belong to the Financial sector occupying the two top positions. Most stocks of the Financial sector are both major senders and major receivers of information.

\scriptsize
\[ \begin{array}{c|l|l|l} \hline \text{\bf In Node Strength} & \text{\bf Company} & \text{\bf Sector} & \text{\bf Industry} \\ \hline 269.38 & \text{Lincoln National Corp} & \text{Financial} & \text{Insurance} \\ 267.27 & \text{Hartford Financial Services Group Inc/The} & \text{Financial} & \text{Insurance} \\ 262.53 & \text{Huntington Bancshares Inc/OH} & \text{Financial} & \text{Banks} \\ 259.25 & \text{Principal Financial Group Inc} & \text{Financial} & \text{Insurance} \\ 254.91 & \text{American International Group Inc} & \text{Financial} & \text{Insurance} \\ 254.70 & \text{Regions Financial Corp} & \text{Financial} & \text{Banks} \\ 253.71 & \text{E*TRADE Financial Corp} & \text{Financial} & \text{Diversified Financial Services} \\ 251.81 & \text{Prudential Financial Inc} & \text{Financial} & \text{Insurance} \\ 250.69 & \text{Fifth Third Bancorp} & \text{Financial} & \text{Banks} \\ 249.77 & \text{Citigroup Inc} & \text{Financial} & \text{Banks} \\ \hline \text{\bf Out Node Strength} & \text{\bf Company} & \text{\bf Industry} & \text{\bf Sub Industry} \\ \hline 288.90 & \text{E*TRADE Financial Corp} & \text{Financial} & \text{Diversified Financial Services} \\ 276.41 & \text{Lincoln National Corp} & \text{Financial} & \text{Insurance} \\ 275.20 & \text{Lennar Corp} & \text{Consumer, Cyclical} & \text{Home Builders} \\ 274.94 & \text{Cliffs Natural Resources Inc} & \text{Basic Materials} & \text{Iron / Steel} \\ 274.57 & \text{Huntington Bancshares Inc/OH} & \text{Financial} & \text{Banks} \\ 273.39 & \text{American International Group Inc} & \text{Financial} & \text{Insurance} \\ 272.25 & \text{Allegheny Technologies Inc} & \text{Basic Materials} & \text{Iron / Steel} \\ 271.55 & \text{Sirius XM Holdings Inc} & \text{Communications} & \text{Media} \\ 269.63 & \text{Hartford Financial Services Group Inc/The} & \text{Financial} & \text{Insurance} \\ 269.44 & \text{Regions Financial Corp} & \text{Financial} & \text{Banks} \\ \hline \end{array} \]

\normalsize

\vskip 0.2 cm

\noindent {\bf Table 3.} Classification of stocks with highest Node Strength, their sector and industry classifications. Only the ten stocks with highest centrality values are shown.

\vskip 0.3 cm

\subsection{Aggregate Data}

In order to understand how sectors relate to one another, we made use of aggregate data, which resulted in correlations and TE relations between sectors. The aggregate data was constructed in the following way: first, we calculated the correlations between stocks belonging to the same sector, and then constructed a correlation matrix for each sector. From those correlation matrices, we calculated eigenvalues and eigenvectors for each correlation matrix, with each eigenvector being a vector associated to one of the scalar eigenvalues. For each of them, one eigenvalue detaches from all others, being much larger than most of them, what was first seen in stock markets in the work of Laloux, Cizeau, Bouchaud, and Potters (1999), and verified by a number of other works since then for many types of markets (for a comprehensive bibliography on the subject, see Sandoval and Franca, 2012).

Each eigenvalue may be seen as an indicator of the level of risk of a portfolio built with the stocks (or indices) of the correlation matrix by using the elements of the eigenvector associated with it as weights for each stock in the portfolio. So, the highest eigenvalue expresses the risk of a portfolio with the maximum risk, and associated to this eigenvalue is an eigenvector that is remarkably homogeneous in terms of the weights given to each element of this portfolio. This largest eigenvalue is then associated with the systemic risk of the market, being the corresponding eigenvector associated with a {\sl market mode}.

The agreement of an index built using as weights the elements of the eigenvector corresponding to the largest eigenvalue of the correlation matrix of stocks of a certain stock market, when compared to that stock market index, as an example, is nearly perfect. So, we may use this eigenvector in order to build an index for each sector, and then use all indices thus obtained in order to calculate a correlation matrix and a TE matrix. The results are plotted in Figure 11. Both graphs reveal similar information, indicating again a relation between the transfer of information of a sector from one day before to the next day of another sector and the correlation of both in the next day. There is a lot of interaction between the Basic Materials sector and the Industrial Sector, and the Energy sector doesn't seem to interact much with the others, except for a light interaction with the Basic Materials sector. The Industrial sector interacts strongly with the Consumer, Cyclical, Financial, Communications and Technology sectors, what also happens in a lesser degree with the interactions of the Financial sector. Actually, these 5 sectors form a highly interacting block, with stronger interactions between the Communications and Technology sectors. It is worthwhile remembering that the one stock belonging to the Diversified sector has been incorporated into the Financial sector.

Tables 4 and 5 show the Node Strengths, and the In and Out Node Strengths, for the correlation and the TE matrices. The most central sector according to correlation is the Industrial one, followed by Communications. The sectors wit the smallest value for Node Strength are Energy and Utilities. According to TE, the Industrial sector is the main sender and receiver of information, followed by the Communications and Financial sectors. The Energy and the Utilities sectors are the ones that send and receive the least information from the other sectors, although it is worth remembering that the energy sector is very much correlated with itself and also share much information among its stocks.

\scriptsize
\[ \begin{array}{c|l} \hline \text{\bf Node Strength} & \text{\bf Sector} \\ \hline 6.80 & \text{Industrial} \\ 6.67 & \text{Communications} \\ 6.49 & \text{Consumer, Non-Cyclical} \\ 6.47 & \text{Basic Materials} \\ 6.38 & \text{Consumer, Cyclical} \\ 6.22 & \text{Technology} \\ 6.08 & \text{Financial} \\ 5.53 & \text{Energy} \\ 5.44 & \text{Utilities} \\ \hline \end{array} \]

\normalsize

\vskip 0.2 cm

\noindent {\bf Table 4.} Classification of industries according to Node Strength for aggregate data on sectors.

\vskip 0.3 cm

\scriptsize
\[ \begin{array}{c|l||c|l} \hline \text{\bf In Node Strength} & \text{\bf Sector} & \text{\bf Out Node Strength} & \text{\bf Industry} \\ \hline 4.61 & \text{Industrial} & 4.62 & \text{Industrial} \\ 4.25 & \text{Communications} & 4.24 & \text{Financial} \\ 4.18 & \text{Financial} & 4.23 & \text{Communications} \\ 4.16 & \text{Basic Materials} & 4.17 & \text{Consumer, Cyclical} \\ 4.13 & \text{Consumer, Cyclical} & 4.12 & \text{Basic Materials} \\ 3.88 & \text{Technology} & 3.88 & \text{Technology} \\ 3.50 & \text{Consumer, Non-Cyclical} & 3.45 & \text{Consumer, Non-Cyclical} \\ 2.94 & \text{Energy} & 3.01 & \text{Energy} \\ 2.42 & \text{Utilities} & 2.34 & \text{Utilities} \\ \hline \end{array} \]

\normalsize

\vskip 0.2 cm

\noindent {\bf Table 5.} Classification of sectors according to In and Out Node Strengths for aggregate data on sectors.

\vskip 0.3 cm

One cans also see in Table 5 that there is a slight asymmetry between the In and Out Node Strengths of sectors. As an example, the Financial sector sends slightly more information than it receives, and the Communications sector receives slightly more information than it sends. It is interesting to analyze these asymmetries between sent and received information, what is done in the next section.

\subsection{Asymmetries in Transfer Entropy}

As we could see, although Transfer Entropy is an asymmetric measure, it is highly symmetric if we consider the Transfer Entropy from lagged variables to original ones. The differences between $T_{I\rightarrow J}$ and $T_{J\rightarrow I}$ will be called here Excess Transfer Entropy, defined in terms of sector 21 of the TE matrix as
\begin{equation}
\label{excessTE}
\text{ExcessTE}(i,j)=\text{TESect21}(i,j)-\text{TESect21}(j,i)\ ,
\end{equation}
where TESect21 is just sector 21 of the TE matrix. The result is an antisymmetric matrix with information on the difference between the amount of information a time series of a stock $I$ transfer to the time series of another stock $J$ on the next day and the amount of information that stock $J$ transfer to stock $I$ of the next day. For our set of data, it ranges from -0.0897 to 0.0897.

The graphs of both the ExcessTE matrix for all stocks and the ExcessTE matrix for aggregate data by sectors are drawn in Figure 12 (left graph for stocks and right graph for sectors). The first graph highlights regions of high excess TE, from some particular stocks to all others, the main region being due to lagged variables to the stocks belonging to the Utilities sector. For aggregate data (right graph), there is an Excess TE from the Energy sector to the Basic Materials, Industrial, Consumer, Cyclical, Financial, Communications, Utilities, and Consumer, Non-Cyclical sectors. There is also a strong Excess TE from Consumer, Cyclical to Basic Materials, and to the Technology, Utilities and Consumer, Non-Cyclical sectors. The Financial sector also has an Excess TE to these same sectors, with emphasis on Excess TE to the Utilities sector.

Table 6 shows the In and Out Node Strengths of the top 10 stocks that have the major imbalances between the information they send and the information they receive from all other stocks. The major Excess receivers are a major oil company and a major pharmaceutical company, followed by two energy companies, and the top excess senders are quite diverse.

In order to analyze which are the major excess sender and receiver sectors, we calculate the In and Out Node Strengths of the Excess TE data for aggregate data. The result is in Table 7, showing the positions of all sectors. The top two senders of Excess TE are the Energy and Financial sectors, and the major Excess TE receivers are the Utilities and Basic Materials sectors.

\scriptsize
\[ \begin{array}{c|l|l|l} \hline \text{\bf In Node Strength} & \text{\bf Company} & \text{\bf Sector} & \text{\bf Industry} \\ \hline 11.42 & \text{Exxon Mobil Corp} & \text{Energy} & \text{Oil  \&  Gas} \\ 10.93 & \text{Johnson \& Johnson} & \text{Consumer, Non-Cyclical} & \text{Pharmaceuticals} \\ 10.70 & \text{Entergy Corp} & \text{Utilities} & \text{Electric} \\ 10.65 & \text{Wisconsin Energy Corp} & \text{Utilities} & \text{Electric} \\ 10.32 & \text{Chevron Corp} & \text{Energy} & \text{Oil  \&  Gas} \\ 10.17 & \text{American Electric Power Co Inc} & \text{Utilities} & \text{Electric} \\ 10.05 & \text{Praxair Inc} & \text{Basic Materials} & \text{Chemicals} \\ 9.92 & \text{PepsiCo Inc} & \text{Consumer, Non-Cyclical} & \text{Beverages} \\ 9.87 & \text{HJ Heinz Co} & \text{Consumer, Non-Cyclical} & \text{Food} \\ 9.87 & \text{Northeast Utilities} & \text{Utilities} & \text{Electric} \\ \hline \text{\bf Out Node Strength} & \text{\bf Company} & \text{\bf Sector} & \text{\bf Industry} \\ \hline 22.42 & \text{Sirius XM Holdings Inc} & \text{Communications} & \text{Media} \\ 22.19 & \text{Titanium Metals Corp} & \text{Basic Materials} & \text{Mining} \\ 21.93 & \text{JDS Uniphase Corp} & \text{Communications} & \text{Telecommunications} \\ 20.26 & \text{Allegheny Technologies Inc} & \text{Basic Materials} & \text{Iron / Steel} \\ 19.83 & \text{ Lennar Corp} & \text{Consumer, Cyclical} & \text{Home Builders} \\ 19.24 & \text{Monster Beverage Corp} & \text{Consumer, Non-Cyclical} & \text{Beverages} \\ 19.09 & \text{Netflix Inc} & \text{Communications} & \text{Internet} \\ 18.66 & \text{Regeneron Pharmaceuticals Inc} & \text{Consumer, Non-Cyclical} & \text{Biotechnology} \\ 17.98 & \text{Advanced Micro Devices Inc} & \text{Technology} & \text{Semiconductors} \\ 17.43 & \text{NVIDIA Corp} & \text{Technology} & \text{Semiconductors} \\ \hline \end{array} \]

\normalsize

\vskip 0.2 cm

\noindent {\bf Table 6.} Classification of stocks with highest In and Out Node Strengths, based on Excess TE, their sector and industry classifications. Only the ten stocks with highest centrality values are shown.

\vskip 0.3 cm

\scriptsize
\[ \begin{array}{c|l||c|l} \hline \text{\bf In Excess Node Strength} & \text{\bf Sector} & \text{\bf Out Excess Node Strength} & \text{\bf Sector} \\ \hline 0.063 & \text{Energy} & 0.077 & \text{Utilities} \\ 0.060 & \text{Financial} & 0.043 & \text{Basic Materials} \\ 0.045 & \text{Consumer, Cyclical} & 0.042 & \text{Consumer, Non-Cyclical} \\ 0.014 & \text{Industrial} & 0.025 & \text{Communications} \\ 0.004 & \text{Technology} & & \\ \hline \end{array} \]

\normalsize

\vskip 0.2 cm

\noindent {\bf Table 7.} Classification of stocks with highest In and Out Node Strengths, based on Excess TE, for aggregate data by sectors. Negative values have not been represented.

\vskip 0.3 cm

\subsection{Asset Graphs}

As mentioned before, there are many ways to filter the large amount of information provided by the correlation and the TE matrices. One way to do so is to use {\sl asset graphs}, which are networks built by establishing a threshold value above or below which all interactions are not considered, leaving fewer connections and nodes. Asset graphs have been used in a variety of works in finance, like in Onela, Chakraborti, Kaski, and Kert\'{e}sz (2002), (2003), (2004), Onela, Chakraborti, Kaski, and Kert\'{e}sz, and Kanto (2003), Onnela, Chakraborti, and Kaski (2003), Sinha and Pan (2003), Ausloos and Lambiotte (2007), and Sandoval (2012).

In our particular case, we shall use threshold values for the correlation matrix and for the TE matrix below which edges between the nodes (stocks) are not considered, and all nodes not connected to any other node are also deleted. We used as examples threshold values that made it possible to distinguish between many clusters, a compromise between having too many connections or too few of them. For correlation, we used a threshold value 0.8. So, a new matrix was built on the correlation matrix, where all elements of the matrix for which the corresponding element in the correlation matrix was below the threshold was set to zero and all elements above it were set to one, what is called an {\sl adjacency matrix}. Then, all nodes without any connection were removed, resulting in a network of fewer nodes and fewer edges. The result is plotted in Figure 13, where one can see 13 clusters, some of them as small as two nodes and the largest of them, corresponding to REITS, with 14 nodes. The sectors and industries that make up the clusters are specified in the same figure, with large networks of Oil \& Gas companies and also of Banks. Stocks are represented by their tickers, as in Appendix A.

An asset graph based on TE with threshold value 0.7 is built and is shown in Figure 14. The clusters formed are similar to the clusters obtained from correlation, but now the cluster of REITS is smaller, the cluster of banks is larger, and there is just one large cluster of Oil \& Gas companies. There is also a cluster of Insurance companies now. In this network, single directed links are represented with arrows, and links that occur in both directions are represented as lines without arrows.

\section{Dynamics}

The use of data for the production of static measurements of the relations between stocks offer some information about them, but the dynamics of those relations reveal how they evolve in time and how they react to external phenomena and to consequences of their own interactions. In this section, we study the dynamics of the correlations and of the exchange of information of the time series of stocks as measured by the Pearson correlation and Transfer Entropy, respectively. We start by considering snapshots by semester, and then use running windows in order to study the development of the relations between stocks as measured by correlation and by Transfer Entropy.

The correlations between stocks change in time, becoming larger in times of crisis. Figure 15 shows correlation matrices calculated using data semester by semester, from 2003 to 2012. It is easy to notice the much brighter colors in the second semester of 2008 (the height of the US Subprime Crisis) and in the second semester of 2011 (the height of the European Sovereign Debt Crisis). The average correlation in 2011 was even higher than in the crisis of 2008, thus showing that during part of the European Sovereign Debt Crisis the stocks of the US market behaved essentially in the same way. Looking at sectors, the Energy sector is very connected in itself, but not very connected to others, particularly prior to 2008. Some similar behavior can be seen for the Utilities sector, during some periods of time before and after 2008.

In a recent work, Buccheri, Marmi, and Mantegna (2013) also studied the correlations within the US stock market, but using industrial indices on a larger span of time than ours, with comparable results when within the same time span we use.

In Figure 16, we analyze the evolution of TE from lagged to original variables in time, using windows comprising data from each semester from 2003 to 2012. For the calculations in Figure 16, we used a binning of size 0.1, since it is more appropriate due to the small sample for each semester and also faster to calculate. One may see a rise in Transfer Entropy during periods of crisis, like at the 2008 Subprime Crisis and the 2011 European Sovereign Debt Crisis, very much like the rise of correlation in periods of crisis. Just prior to the crisis of 2008, we can detect a rise in the TE between stocks of the REITS industry in the Financial sector, in the Communications sector, and in the Technology sector.

For a more continuous analysis of the evolution of correlation and TE along the years, we considered running windows of 100 days of data each, shifting by one day at a time. For each correlation matrix calculated, we calculated the average of the node strengths of all stocks; for each TE matrix, we calculated the average in and out node strengths of all stocks. Then, by dividing these node strengths by the number of stocks, $N=464$, we obtained what we call the Mean Correlation (average of the node strengths divided by $N$), and the Mean In and Out Transfer Entropy (respectively, the in and out node strengths, divided by $N$). In Figure 17, we plot the mean volatility of stocks (as given by the absolute values of their log-returns) together with the mean correlation (calculated from the correlation matrices) and the mean in and out transfer entropy (calculated from the TE matrices). Since the results for correlation and TE are based on windows of 100 days of data, we plot each result at the last day of the window, so that we never consider effects that are in the future of the day to which the result is associated. As a consequence, all averages on node strengths appear as zero in the first 100 days, what does not happen for the mean volatility.

Looking at Figure 17, the mean correlation is large for the periods of crisis, at the end of 2008, at the begining of 2010, and at the end of 2011. Note that the mean correlation between stocks is larger for the period related with the height of the European Sovereign Debt Crisis than it was during the 2008 Subprime Crisis. The mean in and out TEs are almost indistinguishable from each other, and both have the same behavior as the mean correlation between stocks: there is a rise in mean Transfer Entropy in times of crisis. There are, though, some important differences. The first one is that the mean TE in and the mean TE out both present lower results to the crisis in 2010 than the results obtained with correlation, following more closely the rises and falls of the volatility. Now the mean correlation rises with the crisis of 2008, but then remains high for the remaining time. So, although there hasn't been a steady rise in the exchange of information between stocks, there seems to be a steady rise in the correlation between them.

We may wish to analyze the behavior of stocks separately. This is done by considering the node strength of each stock and their in and out node strengths. In Figure 18, we plot the volatility of each stock in the sample in time, calculated as the absolute value of its log-return, the mean correlation, given by the stock's node strength in the correlation matrix, and the mean in and out TE, given by the stock's in and out node strengths, respectively.

In Figure 18, the vertical lines represent the stocks in the sample (464 of them), with the numbers delimiting each sector, and the horizontal lines represent time, as measured in years. The result of each window of 100 days is associated with the last day of the window, so the measures based on moving windows begin to appear on the 100th day of each graph, except for volatility. Note that, due to the use of windows of 100 days in the calculations of the mean correlation and the mean in and out TEs, their images appear smeared when compared with the image for volatility.

Analyzing the volatility graph, one can see three different episodes of crisis in the 10 years spanned by the data. The main one occurs at the end of 2008, and corresponds to the 2008 Subprime Mortgage Crisis of the USA. It follows two smaller periods of high volatility at the beginning of 2008, and the crisis at the end of the year starts in the Financial sector, rapidly propagating to all the other sectors. The crisis subsides only at the end of 2009, and does not affect much the Utilities sector, and slightly less the Consumer, Non-Cyclic sector. The second crisis hits, not with the same strength of the 2008 crisis, by the middle of 2010. At the time, the first signs of real instability of the Eurozone became clear when the Greek government showed no signs of being able to pay the interest on their foreign debt. By the end of 2011, the third wave of high volatility hits, stronger than the one in 2010, but still weaker than the one of 2008, when the crisis had spread to other countries of the Eurozone and there were questions about the efficacy of the policies determined by the major economic players in Europe.

Looking at the plot of the mean correlation, we can see that all stocks behave similarly in time, increasing their mean correlation with the market in times of crisis. One exception can be seen for the stocks of the Energy sector, which exhibit very low mean correlations just prior to the 2008 US Subprime Crisis (dark spot in the middle of the graph), what may be associated with the 2000s energy crisis. The correlations within the Energy sector remained high for a longer period than the correlation within other industries after the crisis of 2008. Since 2003, the price of the oil barrel (in US dollars) had been growing, and by July 2008, the price spiked, falling strongly after the Subprime Mortgage Crisis. Some reasons suggested for the growing trend before 2008 were the fall of the value of the dollar with respect to other currencies and the excess of demand for energy in a world where various economies were growing fast, together with a fear of limited supply by producers.

Both figures for In Transfer Entropy and Out Transfer Entropy are very similar, what implies that the Transfer Entropy matrix from lagged stocks to original ones (Sector 21) is rather symmetric, what can be also seen from the apparent symmetry of Figure 6. The ${TE}_{in}$ ranges from $0.0127$ to $0.5298$ and the ${TE}_{out}$ ranges from $0.0138$ to $0.4535$. So, the minimum and maximum values of ${TE}_{out}$ are slightly smaller than the minimum and maximum values of ${TE}_{in}$. Again, the mean in and out TEs are weaker than the mean correlations in some periods of crises. In the crisis of 2008, both TE and correlation registered peaks, but at the end of 2010, there was an increase in both volatility and correlation without a similar increase in TE. For the end of 2011, there was high correlation and medium volatility and TE.

\section{Simulations with the network based on Transfer Entropy}

As stated in the introduction to this article, there is a belief that the network structure of banks influence the way a shock may propagate. Usually, banks that are more connected are the major spreaders of shocks, but a bank that is not so connected, but whose connections are themselves central, can also be the agent of the propagation of a shock. There is also the issue of back-reaction: a shock may influence other banks that, on their turn, may influence back the bank from which the shock started, and one may even obtain self reinforcement, so that a small crisis may trigger a much bigger one.

Here, we are using stock prices (log-returns) in order to study how one stock relates with another. So, we are not analyzing just defaults or just the banking system, but a larger system made of stocks of a diversity of industries of the US economy. In our modeling of the propagation of crises, we shall consider the trigger of a shock as a considerable drop on the value of a stock, like a 30\% fall in its value. This shock then propagates on the network as a factor of the original shock being applied to its neighbors (all stocks, in our case), and we shall consider this factor to be proportional to the TE from the stock to each of its neighbors. We shall be using the network built on TE and not on correlation because we want the shocks to propagate from one day to another (we shall assume a day as our unit of time). In order to avoid influence from one stock to itself on the next day, what would lead to self reinforcement of effects, we set the main diagonal elements of the TE matrix for Sector 21 to zero.

Also, in order to make the effect of shocks decrease in time, we will multiply each iteration by a negative exponential. Our model for the propagation of shocks may be described by the following equation:
\begin{equation}
\label{model}
V_{i,t+1}=\sum_{j=1}^NV_{j,t}^TMTE_{ij}{\ \rm e}^{-(t+1)}\ ,
\end{equation}
where $V_{j,t}^T$ is the transpose of the volatility (absolute value of the log-return) of stock $j$ at time $t$, $V_{i,t+1}$ is the volatility of stock $i$ at time $t+1$ (units of one day), and $MTE_{ij}$ is the Transfer Entropy from $i$ to $j$ (from lagged to original variables) with the main diagonal set to zero. So, the volatility of a stock on the following day will depend on the sum of the transfer entropies of all other stocks, multiplied by the volatilities of those stocks on the previous day, times a decreasing exponential factor.

This is a very crude model, which does not take into account a diversity of factors, like the reaction of prices when they have a fall or rise that is beyond the value that the market sees fit for a stock, or the capacity of some stocks of absorbing shocks, but it gives some interesting enough results for the propagation of shocks in a real network of stocks.

In order to use our model for the observation of the propagation of shocks, we shall start with all volatilities set to zero and set one of them as $0.3$, which would be equivalent to a large fall (or rise) of the stock of the company. Then, we calculate the volatilities of all stocks, including the original one, by using (\ref{model}). The results are then used in the calculation of the next period of time and so on, until the shock has subsided. We do this for all stocks in the sample, each one starting its own shock, and analyze the differences in the propagation of the shocks.

Figure 19 shows the effect of a shock starting with the J. P. Morgan (bank), with Exxon Mobil (oil \& gas), and with Microsoft (computer software). The graphs show a three dimensional view of the volatility in time according to stock. We can see that the shocks propagate very rapidly to all stocks, but they hit different sectors with distinct strengths. According to the simulations, a high volatility in the stocks of Exxon Mobil (oil \& gas) would cause a greater shock than a similar volatility in the stocks of the J. P. Morgan, and a high volatility in the stocks of Microsoft (computer software) would lead to an average propagation when comparing to the two other stocks.

Figure 20 shows the propagation of a shock to the J. P. Morgan bank on the network built using correlation, which is very similar to the network obtained with TE, but which is built without the need for manually imposing constraints. Higher volatility is represented by darker dots and lower volatility is represented by brighter dots. The shock starting at the Financial sector spreads to all sectors, with low intensity first, and then causes an increasing wave of volatility that concentrates on the densely populated region involving most sectors, but mainly the Industrial, Communications and Technology sectors, fading away after that. Although our model is rather crude, it replicates the behavior that can be observed from the volatilities in Figure 11, where a rise in one particular sector (Financial, in that case) rapidly propagates to all other stocks.

As the shocks reach their heights at day 4, we shall use the average volatility at this day as a measure of how strong is the wave produced by the shock in one stock. We shall call this measure Shock Propagation Strength. Table 8 shows the top 10 stocks according to this measure, their company names, sectors, and industries. Most belong to the Technology and Communications sectors, which are some of the most central in the network. So, although they do not have the largest Node Strengths, or In and Out Node Strengths, they are reasonably central in a region of central nodes. Belonging to a more populated area of the network, they are the most likely ones to propagate shocks when we consider our crude model.

\scriptsize
\[ \begin{array}{c|l|l|l} \hline \text{\bf Node Strength} & \text{\bf Company} & \text{\bf Sector} & \text{\bf Industry} \\ \hline 0.217 & \text{Hewlett-Packard Co} & \text{Technology} & \text{Computers} \\ 0.214 & \text{Corning Inc} & \text{Communications} & \text{Telecommunications} \\ 0.210 & \text{Cliffs Natural Resources Inc} & \text{Basic Materials} & \text{Iron/Steel} \\ 0.209 & \text{Western Digital Corp} & \text{Technology} & \text{Computers} \\ 0.208 & \text{Dell Inc} & \text{Technology} & \text{Computers} \\ 0.206 & \text{Frontier Communications Corp} & \text{Communications} & \text{Telecommunications} \\ 0.205 & \text{Symantec Corp} & \text{Communications} & \text{Internet} \\ 0.205 & \text{Sirius XM Holdings Inc} & \text{Communications} & \text{Telecommunications} \\ 0.204 & \text{Amphenol Corp} & \text{Industrial} & \text{Electronics} \\ 0.203 & \text{Nucor Corp} & \text{Basic Materials} & \text{Iron/Steel} \\ \hline \end{array} \]

\normalsize

\vskip 0.2 cm

\noindent {\bf Table 8.} Classification of stocks with highest Shock Propagation Strength, their sector and industry classifications. Only the ten stocks with highest values are shown.

\vskip 0.3 cm

We may also apply a systemic shock to all stocks by setting, for example, all volatilities to 0.1, or apply shocks to sectors. Figure 21 shows the effects of shocks of intensity 0.1 to, respectively, all stocks, stocks of the Financial sector, and stocks of the Technology sector. Due to the rapid spread of shocks in the network and to the model used, all figures look similar, expect for scale, that is much higher if the shock is applied to all stocks, and similar in the case of the shock being applied to the Financial or to the Technology sectors.

Because of the way the simulation was constructed, stocks that are more influential in terms of having high Out Node Strengths, and which ``live'' in regions where stocks are also more connected are the most likely ones to disseminate a crisis. Other types of models would lead to different results, but they closely follow other results based on other types of network models which only include banks or other financial institutions.

In order to close this article, we would like to show some figures that depict the Node Strengths and Shock Propagation Strength on the network formed by correlations, all represent in Figure 20. The dots represent the positions occupied by stocks in the network, which is the same type of graph obtained in Figure 9. Darker colors represent higher values of the measures being depicted and brighter colors represent lower values for these measures. The first graph shows the Node Strength, derived from correlations, and the second and third graphs show the In and Out Node Strengths derived from Transfer Entropy. The last graph shows the Shock Propagation Strength, calculated in this section. Node Strength spreads mostly evenly among stocks, and In and Out Node Strengths tend to concentrate in the Financial sector. Now, the Shock Propagation Strength concentrates in regions of high concentrations of nodes.

\section{Conclusions}

The two networks built using the correlations between stocks of the US stock market and using the Transfer Entropy from stocks in one day to stocks on the next day proved to be very similar, indicating that a large exchange of information between stocks from one day to another is associated with their similar behavior on this next day. We could see the importance of some sectors of the US economy in the dissemination of information, like the Financial and the Industrial sectors, and the role of receivers of information like the Utilities and Energy sectors. By using moving windows, we could see how correlation and Transfer Entropy rise in times of crises and how correlation has been growing after the crisis of 2008, what did not happen to Transfer Entropy. By building a model based on Transfer Entropy, we could simulate how volatility may propagate in a network of stocks and how sectors that occupy more central positions, like the Communications and Technology sectors, have major roles in the propagation of crises. Some future work will involve working with high frequency data and developing better models for the spread of crises.

\vskip 0.6 cm

\noindent{\bf Acknowledgements}

\vskip 0.4 cm

This article was written using \LaTeX, all figures were made using PSTricks and Matlab, and the calculations were made using Matlab and Excel.

\newpage

\appendix

\section{List of stocks}

\tiny

\[ \! \! \! \! \! \! \! \! \! \! \begin{array}{c|c|l|l|l|l} \\ & \text{Ticker} & \text{Company} & \text{Sector} & \text{Industry} & \text{Sub-Industry}
\\ \hline 1 & \text{MON} & \text{Monsanto Co (MON)} & \text{Basic Materials} & \text{Chemicals} & \text{Agricultural Chemicals}
\\ 2 & \text{MOS} & \text{Mosaic Co/The (MOS)} & \text{Basic Materials} & \text{Chemicals} & \text{Agricultural Chemicals}
\\ 3 & \text{DOW} & \text{Dow Chemical Co/The (DOW)} & \text{Basic Materials} & \text{Chemicals} & \text{Chemicals-Diversified}
\\ 4 & \text{DD} & \text{EI du Pont de Nemours \& Co (DD)} & \text{Basic Materials} & \text{Chemicals} & \text{Chemicals-Diversified}
\\ 5 & \text{FMC} & \text{FMC Corp (FMC)} & \text{Basic Materials} & \text{Chemicals} & \text{Chemicals-Diversified}
\\ 6 & \text{PPG} & \text{PPG Industries Inc (PPG)} & \text{Basic Materials} & \text{Chemicals} & \text{Chemicals-Diversified}
\\ 7 & \text{EMN} & \text{Eastman Chemical Co (EMN)} & \text{Basic Materials} & \text{Chemicals} & \text{Chemicals-Specialty}
\\ 8 & \text{ECL} & \text{Ecolab Inc (ECL)} & \text{Basic Materials} & \text{Chemicals} & \text{Chemicals-Specialty}
\\ 9 & \text{IFF} & \text{International Flavors \& Fragrances Inc (IFF)} & \text{Basic Materials} & \text{Chemicals} & \text{Chemicals-Specialty}
\\ 10 & \text{SIAL} & \text{Sigma-Aldrich Corp (SIAL)} & \text{Basic Materials} & \text{Chemicals} & \text{Chemicals-Specialty}
\\ 11 & \text{SHW} & \text{Sherwin-Williams Co/The (SHW)} & \text{Basic Materials} & \text{Chemicals} & \text{Coatings/Paint}
\\ 12 & \text{APD} & \text{Air Products \& Chemicals Inc (APD)} & \text{Basic Materials} & \text{Chemicals} & \text{Industrial Gases}
\\ 13 & \text{ARG} & \text{Airgas Inc (ARG)} & \text{Basic Materials} & \text{Chemicals} & \text{Industrial Gases}
\\ 14 & \text{PX} & \text{Praxair Inc (PX)} & \text{Basic Materials} & \text{Chemicals} & \text{Industrial Gases}
\\ 15 & \text{IP} & \text{International Paper Co (IP)} & \text{Basic Materials} & \text{Forest Products \& Paper} & \text{Paper \& Related Products}
\\ 16 & \text{MWV} & \text{MeadWestvaco Corp (MWV)} & \text{Basic Materials} & \text{Forest Products \& Paper} & \text{Paper \& Related Products}
\\ 17 & \text{CLF} & \text{Cliffs Natural Resources Inc (CLF)} & \text{Basic Materials} & \text{Iron / Steel} & \text{Metal-Iron}
\\ 18 & \text{NUE} & \text{Nucor Corp (NUE)} & \text{Basic Materials} & \text{Iron / Steel} & \text{Steel-Producers}
\\ 19 & \text{X} & \text{United States Steel Corp (X)} & \text{Basic Materials} & \text{Iron / Steel} & \text{Steel-Producers}
\\ 20 & \text{TIE} & \text{Titanium Metals Corp (TIE)} & \text{Basic Materials} & \text{Iron / Steel} & \text{Steel-Producers}
\\ 21 & \text{ATI} & \text{Allegheny Technologies Inc (ATI)} & \text{Basic Materials} & \text{Iron / Steel} & \text{Steel-Specialty}
\\ 22 & \text{NEM} & \text{Newmont Mining Corp (NEM)} & \text{Basic Materials} & \text{Mining} & \text{Gold Mining}
\\ 23 & \text{GOLD} & \text{Randgold Resources Ltd (GOLD)} & \text{Basic Materials} & \text{Mining} & \text{Gold Mining}
\\ 24 & \text{AA} & \text{Alcoa Inc (AA)} & \text{Basic Materials} & \text{Mining} & \text{Metal-Aluminum}
\\ 25 & \text{FCX} & \text{Freeport-McMoRan Copper \& Gold Inc (FCX)} & \text{Basic Materials} & \text{Mining} & \text{Metal-Copper}
\\ 26 & \text{VMC} & \text{Vulcan Materials Co (VMC)} & \text{Basic Materials} & \text{Mining} & \text{Quarrying}
\\ 27 & \text{CNX} & \text{CONSOL Energy Inc (CNX)} & \text{Energy} & \text{Coal} & \text{Coal}
\\ 28 & \text{BTU} & \text{Peabody Energy Corp (BTU)} & \text{Energy} & \text{Coal} & \text{Coal}
\\ 29 & \text{APC} & \text{Anadarko Petroleum Corp (APC)} & \text{Energy} & \text{Oil \& Gas} & \text{Oil Comp-Explor \& Prodtn}
\\ 30 & \text{APA} & \text{Apache Corp (APA)} & \text{Energy} & \text{Oil \& Gas} & \text{Oil Comp-Explor \& Prodtn}
\\ 31 & \text{COG} & \text{Cabot Oil \& Gas Corp (COG)} & \text{Energy} & \text{Oil \& Gas} & \text{Oil Comp-Explor \& Prodtn}
\\ 32 & \text{CHK} & \text{Chesapeake Energy Corp (CHK)} & \text{Energy} & \text{Oil \& Gas} & \text{Oil Comp-Explor \& Prodtn}
\\ 33 & \text{DNR} & \text{Denbury Resources Inc (DNR)} & \text{Energy} & \text{Oil \& Gas} & \text{Oil Comp-Explor \& Prodtn}
\\ 34 & \text{DVN} & \text{Devon Energy Corp (DVN)} & \text{Energy} & \text{Oil \& Gas} & \text{Oil Comp-Explor \& Prodtn}
\\ 35 & \text{EOG} & \text{EOG Resources Inc (EOG)} & \text{Energy} & \text{Oil \& Gas} & \text{Oil Comp-Explor \& Prodtn}
\\ 36 & \text{NFX} & \text{Newfield Exploration Co (NFX)} & \text{Energy} & \text{Oil \& Gas} & \text{Oil Comp-Explor \& Prodtn}
\\ 37 & \text{NBL} & \text{Noble Energy Inc (NBL)} & \text{Energy} & \text{Oil \& Gas} & \text{Oil Comp-Explor \& Prodtn}
\\ 38 & \text{OXY} & \text{Occidental Petroleum Corp (OXY)} & \text{Energy} & \text{Oil \& Gas} & \text{Oil Comp-Explor \& Prodtn}
\\ 39 & \text{PXD} & \text{Pioneer Natural Resources Co (PXD)} & \text{Energy} & \text{Oil \& Gas} & \text{Oil Comp-Explor \& Prodtn}
\\ 40 & \text{RRC} & \text{Range Resources Corp (RRC)} & \text{Energy} & \text{Oil \& Gas} & \text{Oil Comp-Explor \& Prodtn}
\\ 41 & \text{SWN} & \text{Southwestern Energy Co (SWN)} & \text{Energy} & \text{Oil \& Gas} & \text{Oil Comp-Explor \& Prodtn}
\\ 42 & \text{EQT} & \text{EQT Corp (EQT)} & \text{Energy} & \text{Oil \& Gas} & \text{Oil Comp-Explor \& Prodtn}
\\ 43 & \text{CVX} & \text{Chevron Corp (CVX)} & \text{Energy} & \text{Oil \& Gas} & \text{Oil Comp-Integrated}
\\ 44 & \text{COP} & \text{ConocoPhillips (COP)} & \text{Energy} & \text{Oil \& Gas} & \text{Oil Comp-Integrated}
\\ 45 & \text{XOM} & \text{Exxon Mobil Corp (XOM)} & \text{Energy} & \text{Oil \& Gas} & \text{Oil Comp-Integrated}
\\ 46 & \text{HES} & \text{Hess Corp (HES)} & \text{Energy} & \text{Oil \& Gas} & \text{Oil Comp-Integrated}
\\ 47 & \text{MRO} & \text{Marathon Oil Corp (MRO)} & \text{Energy} & \text{Oil \& Gas} & \text{Oil Comp-Integrated}
\\ 48 & \text{MUR} & \text{Murphy Oil Corp (MUR)} & \text{Energy} & \text{Oil \& Gas} & \text{Oil Comp-Integrated}
\\ 49 & \text{TSO} & \text{Tesoro Corp (TSO)} & \text{Energy} & \text{Oil \& Gas} & \text{Oil Refining \& Marketing}
\\ 50 & \text{VLO} & \text{Valero Energy Corp (VLO)} & \text{Energy} & \text{Oil \& Gas} & \text{Oil Refining \& Marketing}
\\ 51 & \text{DO} & \text{Diamond Offshore Drilling Inc (DO)} & \text{Energy} & \text{Oil \& Gas} & \text{Oil \& Gas Drilling}
\\ 52 & \text{HP} & \text{Helmerich \& Payne Inc (HP)} & \text{Energy} & \text{Oil \& Gas} & \text{Oil \& Gas Drilling}
\\ 53 & \text{NBR} & \text{Nabors Industries Ltd (NBR)} & \text{Energy} & \text{Oil \& Gas} & \text{Oil \& Gas Drilling}
\\ 54 & \text{NE} & \text{Noble Corp (NE)} & \text{Energy} & \text{Oil \& Gas} & \text{Oil \& Gas Drilling}
\\ 55 & \text{RDC} & \text{Rowan Cos Plc (RDC)} & \text{Energy} & \text{Oil \& Gas} & \text{Oil \& Gas Drilling}
\\ 56 & \text{CAM} & \text{Cameron International Corp (CAM)} & \text{Energy} & \text{Oil \& Gas Services} & \text{Oil Field Mach \& Equip}
\\ 57 & \text{FTI} & \text{FMC Technologies Inc (FTI)} & \text{Energy} & \text{Oil \& Gas Services} & \text{Oil Field Mach \& Equip}
\\ 58 & \text{NOV} & \text{National Oilwell Varco Inc (NOV)} & \text{Energy} & \text{Oil \& Gas Services} & \text{Oil Field Mach \& Equip}
\\ 59 & \text{BHI} & \text{Baker Hughes Inc (BHI)} & \text{Energy} & \text{Oil \& Gas Services} & \text{Oil-Field Services}
\\ 60 & \text{HAL} & \text{Halliburton Co (HAL)} & \text{Energy} & \text{Oil \& Gas Services} & \text{Oil-Field Services}
\\ 61 & \text{SLB} & \text{Schlumberger Ltd (SLB)} & \text{Energy} & \text{Oil \& Gas Services} & \text{Oil-Field Services}
\\ 62 & \text{WMB} & \text{Williams Cos Inc/The (WMB)} & \text{Energy} & \text{Pipelines} & \text{Pipelines}
\\ 63 & \text{OKE} & \text{ONEOK Inc (OKE)} & \text{Energy} & \text{Pipelines} & \text{Pipelines}
\\ 64 & \text{BA} & \text{Boeing Co/The (BA)} & \text{Industrial} & \text{Aerospace/Defense } & \text{Aerospace/Defense}
\\ 65 & \text{GD} & \text{General Dynamics Corp (GD)} & \text{Industrial} & \text{Aerospace/Defense } & \text{Aerospace/Defense}
\\ 66 & \text{LMT} & \text{Lockheed Martin Corp (LMT)} & \text{Industrial} & \text{Aerospace/Defense } & \text{Aerospace/Defense}
\\ 67 & \text{NOC} & \text{Northrop Grumman Corp (NOC)} & \text{Industrial} & \text{Aerospace/Defense } & \text{Aerospace/Defense}
\\ 68 & \text{RTN} & \text{Raytheon Co (RTN)} & \text{Industrial} & \text{Aerospace/Defense } & \text{Aerospace/Defense}
\\ 69 & \text{COL} & \text{Rockwell Collins Inc (COL)} & \text{Industrial} & \text{Aerospace/Defense } & \text{Aerospace/Defense}
\\ 70 & \text{UTX} & \text{United Technologies Corp (UTX)} & \text{Industrial} & \text{Aerospace/Defense } & \text{Aerospace/Defense-Equip}
\\ 71 & \text{LLL} & \text{L 3 Communications Holdings Inc (LLL)} & \text{Industrial} & \text{Aerospace / Defense} & \text{Electronics-Military}
\\ 72 & \text{MAS} & \text{Masco Corp (MAS)} & \text{Industrial} & \text{Building Materials} & \text{Bldg Prod-Wood}
\\ 73 & \text{EMR} & \text{Emerson Electric Co (EMR)} & \text{Industrial} & \text{Electrical Comp. \& Equip.} & \text{Electric Products-Misc}
\\ 74 & \text{MOLX} & \text{Molex Inc (MOLX)} & \text{Industrial} & \text{Electrical Comp. \& Equip.} & \text{Electric Products-Misc}
\\ 75 & \text{JBL} & \text{Jabil Circuit Inc (JBL)} & \text{Industrial} & \text{Electronics} & \text{Electronic Compo-Misc}
\\ 76 & \text{GRMN} & \text{Garmin Ltd (GRMN)} & \text{Industrial} & \text{Electronics} & \text{Electronic Compo-Misc}
\\ 77 & \text{APH} & \text{Amphenol Corp (APH)} & \text{Industrial} & \text{Electronics} & \text{Electronic Connectors}
\\ 78 & \text{A} & \text{Agilent Technologies Inc (A)} & \text{Industrial} & \text{Electronics} & \text{Electronic Measur Instr}
\\ 79 & \text{FLIR} & \text{FLIR Systems Inc (FLIR)} & \text{Industrial} & \text{Electronics} & \text{Electronic Measur Instr}
\\ 80 & \text{TYC} & \text{Tyco International Ltd (TYC)} & \text{Industrial} & \text{Electronics} & \text{Electronic Secur Devices}
\\ 81 & \text{HON} & \text{Honeywell International Inc (HON)} & \text{Industrial} & \text{Electronics} & \text{Instruments-Controls}
\\ 82 & \text{PKI} & \text{PerkinElmer Inc (PKI)} & \text{Industrial} & \text{Electronics} & \text{Instruments-Scientific}
\\ 83 & \text{TMO} & \text{Thermo Fisher Scientific Inc (TMO)} & \text{Industrial} & \text{Electronics} & \text{Instruments-Scientific}
\\ 84 & \text{WAT} & \text{Waters Corp (WAT)} & \text{Industrial} & \text{Electronics} & \text{Instruments-Scientific}
\\ 85 & \text{FLR} & \text{Fluor Corp (FLR)} & \text{Industrial} & \text{Engineering \& Construction} & \text{Engineering/R \& D Services}
\\ 86 & \text{JEC} & \text{Jacobs Engineering Group Inc (JEC)} & \text{Industrial} & \text{Engineering \& Construction} & \text{Engineering/R \& D Services}
\\ 87 & \text{SRCL} & \text{Stericycle Inc (SRCL)} & \text{Industrial} & \text{Environmental Control} & \text{Hazardous Waste Disposal}
\\ 88 & \text{RSG} & \text{Republic Services Inc (RSG)} & \text{Industrial} & \text{Environmental Control} & \text{Non-hazardous Waste Disp}
\\ 89 & \text{WM} & \text{Waste Management Inc (WM)} & \text{Industrial} & \text{Environmental Control} & \text{Non-hazardous Waste Disp}
\\ 90 & \text{SNA} & \text{Snap-on Inc (SNA)} & \text{Industrial} & \text{Hand / Machine Tools} & \text{Tools-Hand Held}
\\ 91 & \text{SWK} & \text{Stanley Black \& Decker Inc (SWK)} & \text{Industrial} & \text{Hand / Machine Tools} & \text{Tools-Hand Held}
\\ 92 & \text{CAT} & \text{Caterpillar Inc (CAT)} & \text{Industrial} & \text{Machinery - Constr. \& Min.} & \text{Machinery-Constr \& Mining}
\\ 93 & \text{JOY} & \text{Joy Global Inc (JOY)} & \text{Industrial} & \text{Machinery - Constr. \& Min.} & \text{Machinery-Constr \& Mining}
\\ 94 & \text{CMI} & \text{Cummins Inc (CMI)} & \text{Industrial} & \text{Machinery - Diversified} & \text{Engines-Internal Combust}
\\ 95 & \text{ROK} & \text{Rockwell Automation Inc (ROK)} & \text{Industrial} & \text{Machinery - Diversified} & \text{Industrial Automat/Robot}
\\ 96 & \text{DE} & \text{Deere \& Co (DE)} & \text{Industrial} & \text{Machinery - Diversified} & \text{Machinery-Farm}
\\ 97 & \text{ROP} & \text{Roper Industries Inc (ROP)} & \text{Industrial} & \text{Machinery - Diversified} & \text{Machinery-General Indust}
\end{array} \]

\[ \! \! \! \! \! \! \! \! \! \! \begin{array}{c|c|l|l|l|l} \\ & \text{Ticker} & \text{Company} & \text{Sector} & \text{Industry} & \text{Sub-Industry}
\\ \hline 98 & \text{FLS} & \text{Flowserve Corp (FLS)} & \text{Industrial} & \text{Machinery - Diversified} & \text{Machinery-Pumps}
\\ 99 & \text{PCP} & \text{Precision Castparts Corp (PCP)} & \text{Industrial} & \text{Metal Fabricate / Hardware} & \text{Metal Processors \& Fabrica}
\\ 100 & \text{LEG} & \text{Leggett \& Platt Inc (LEG)} & \text{Industrial} & \text{Miscellaneous Manufacturing} & \text{Diversified Manufact Op}
\\ 101 & \text{MMM} & \text{3M Co (MMM)} & \text{Industrial} & \text{Miscellaneous Manufacturing} & \text{Diversified Manufact Op}
\\ 102 & \text{DHR} & \text{Danaher Corp (DHR)} & \text{Industrial} & \text{Miscellaneous Manufacturing} & \text{Diversified Manufact Op}
\\ 103 & \text{DOV} & \text{Dover Corp (DOV)} & \text{Industrial} & \text{Miscellaneous Manufacturing} & \text{Diversified Manufact Op}
\\ 104 & \text{ETN} & \text{Eaton Corp PLC (ETN)} & \text{Industrial} & \text{Miscellaneous Manufacturing} & \text{Diversified Manufact Op}
\\ 105 & \text{GE} & \text{General Electric Co (GE)} & \text{Industrial} & \text{Miscellaneous Manufacturing} & \text{Diversified Manufact Op}
\\ 106 & \text{ITW} & \text{Illinois Tool Works Inc (ITW)} & \text{Industrial} & \text{Miscellaneous Manufacturing} & \text{Diversified Manufact Op}
\\ 107 & \text{IR} & \text{Ingersoll-Rand PLC (IR)} & \text{Industrial} & \text{Miscellaneous Manufacturing} & \text{Diversified Manufact Op}
\\ 108 & \text{PH} & \text{Parker Hannifin Corp (PH)} & \text{Industrial} & \text{Miscellaneous Manufacturing} & \text{Diversified Manufact Op}
\\ 109 & \text{TXT} & \text{Textron Inc (TXT)} & \text{Industrial} & \text{Miscellaneous Manufacturing} & \text{Diversified Manufact Op}
\\ 110 & \text{PLL} & \text{Pall Corp (PLL)} & \text{Industrial} & \text{Miscellaneous Manufacturing} & \text{Filtration/Separat Prod}
\\ 111 & \text{BLL} & \text{Ball Corp (BLL)} & \text{Industrial} & \text{Packaging \& Containers} & \text{Containers-Metal/Glass}
\\ 112 & \text{OI} & \text{Owens-Illinois Inc (OI)} & \text{Industrial} & \text{Packaging \& Containers} & \text{Containers-Metal/Glass}
\\ 113 & \text{BMS} & \text{Bemis Co Inc (BMS)} & \text{Industrial} & \text{Packaging \& Containers} & \text{Containers-Paper/Plastic}
\\ 114 & \text{SEE} & \text{Sealed Air Corp (SEE)} & \text{Industrial} & \text{Packaging \& Containers} & \text{Containers-Paper/Plastic}
\\ 115 & \text{CSX} & \text{CSX Corp (CSX)} & \text{Industrial} & \text{Transportation} & \text{Transport-Rail}
\\ 116 & \text{NSC} & \text{Norfolk Southern Corp (NSC)} & \text{Industrial} & \text{Transportation} & \text{Transport-Rail}
\\ 117 & \text{UNP} & \text{Union Pacific Corp (UNP)} & \text{Industrial} & \text{Transportation} & \text{Transport-Rail}
\\ 118 & \text{CHRW} & \text{CH Robinson Worldwide Inc (CHRW)} & \text{Industrial} & \text{Transportation} & \text{Transport-Services}
\\ 119 & \text{EXPD} & \text{Expeditors International of Washington Inc (EXPD)} & \text{Industrial} & \text{Transportation} & \text{Transport-Services}
\\ 120 & \text{FDX} & \text{FedEx Corp (FDX)} & \text{Industrial} & \text{Transportation} & \text{Transport-Services}
\\ 121 & \text{R} & \text{Ryder System Inc (R)} & \text{Industrial} & \text{Transportation} & \text{Transport-Services}
\\ 122 & \text{UPS} & \text{United Parcel Service Inc (UPS)} & \text{Industrial} & \text{Transportation} & \text{Transport-Services}
\\ 123 & \text{LUV} & \text{Southwest Airlines Co (LUV)} & \text{Consumer, Cyclical} & \text{Airlines} & \text{Airlines}
\\ 124 & \text{COH} & \text{Coach Inc (COH)} & \text{Consumer, Cyclical} & \text{Apparel} & \text{Apparel Manufacturers}
\\ 125 & \text{RL} & \text{Ralph Lauren Corp (RL)} & \text{Consumer, Cyclical} & \text{Apparel} & \text{Apparel Manufacturers}
\\ 126 & \text{VFC} & \text{VF Corp (VFC)} & \text{Consumer, Cyclical} & \text{Apparel} & \text{Apparel Manufacturers}
\\ 127 & \text{NKE} & \text{NIKE Inc (NKE)} & \text{Consumer, Cyclical} & \text{Apparel} & \text{Athletic Footwear}
\\ 128 & \text{F} & \text{Ford Motor Co (F)} & \text{Consumer, Cyclical} & \text{Auto Manufacturers} & \text{Auto-Cars/Light Trucks}
\\ 129 & \text{PCAR} & \text{PACCAR Inc (PCAR)} & \text{Consumer, Cyclical} & \text{Auto Manufacturers} & \text{Auto-Med \& Heavy Duty Trks}
\\ 130 & \text{BWA} & \text{BorgWarner Inc (BWA)} & \text{Consumer, Cyclical} & \text{Auto Parts \& Equipment} & \text{Auto/Trk Prts \& Equip-Orig}
\\ 131 & \text{JCI} & \text{Johnson Controls Inc (JCI)} & \text{Consumer, Cyclical} & \text{Auto Parts \& Equipment} & \text{Auto/Trk Prts \& Equip-Orig}
\\ 132 & \text{GT} & \text{Goodyear Tire \& Rubber Co/The (GT)} & \text{Consumer, Cyclical} & \text{Auto Parts \& Equipment} & \text{Rubber-Tires}
\\ 133 & \text{GPC} & \text{Genuine Parts Co (GPC)} & \text{Consumer, Cyclical} & \text{Distribution/Wholesale} & \text{Distribution/Wholesale}
\\ 134 & \text{FAST} & \text{Fastenal Co (FAST)} & \text{Consumer, Cyclical} & \text{Distribution/Wholesale} & \text{Distribution/Wholesale}
\\ 135 & \text{GWW} & \text{WW Grainger Inc (GWW)} & \text{Consumer, Cyclical} & \text{Distribution/Wholesale} & \text{Distribution/Wholesale}
\\ 136 & \text{FOSL} & \text{Fossil Inc (FOSL)} & \text{Consumer, Cyclical} & \text{Distribution/Wholesale} & \text{Distribution/Wholesale}
\\ 137 & \text{IGT} & \text{International Game Technology (IGT)} & \text{Consumer, Cyclical} & \text{Entertainement} & \text{Casino Services}
\\ 138 & \text{DHI} & \text{DR Horton Inc (DHI)} & \text{Consumer, Cyclical} & \text{Home Builders} & \text{Bldg-Residential/Commer}
\\ 139 & \text{LEN} & \text{Lennar Corp (LEN)} & \text{Consumer, Cyclical} & \text{Home Builders} & \text{Bldg-Residential/Commer}
\\ 140 & \text{PHM} & \text{PulteGroup Inc (PHM)} & \text{Consumer, Cyclical} & \text{Home Builders} & \text{Bldg-Residential/Commer}
\\ 141 & \text{WHR} & \text{Whirlpool Corp (WHR)} & \text{Consumer, Cyclical} & \text{Home Furnishings} & \text{Appliances}
\\ 142 & \text{HAR} & \text{Harman International Industries Inc (HAR)} & \text{Consumer, Cyclical} & \text{Home Furnishings} & \text{Audio/Video Products}
\\ 143 & \text{NWL} & \text{Newell Rubbermaid Inc (NWL)} & \text{Consumer, Cyclical} & \text{Housewares} & \text{Home Decoration Products}
\\ 144 & \text{CCL} & \text{Carnival Corp (CCL)} & \text{Consumer, Cyclical} & \text{Leisure Time} & \text{Cruise Lines}
\\ 145 & \text{HOG} & \text{Harley-Davidson Inc (HOG)} & \text{Consumer, Cyclical} & \text{Leisure Time} & \text{Motorcycle/Motor Scooter}
\\ 146 & \text{WYNN} & \text{Wynn Resorts Ltd (WYNN)} & \text{Consumer, Cyclical} & \text{Lodging} & \text{Casino Hotels}
\\ 147 & \text{MAR} & \text{Marriott International Inc/DE (MAR)} & \text{Consumer, Cyclical} & \text{Lodging} & \text{Hotels \& Motels}
\\ 148 & \text{HOT} & \text{Starwood Hotels \& Resorts Worldwide Inc (HOT)} & \text{Consumer, Cyclical} & \text{Lodging} & \text{Hotels \& Motels}
\\ 149 & \text{ANF} & \text{Abercrombie \& Fitch Co (ANF)} & \text{Consumer, Cyclical} & \text{Retail} & \text{Retail-Apparel/Shoe}
\\ 150 & \text{GPS} & \text{Gap Inc/The (GPS)} & \text{Consumer, Cyclical} & \text{Retail} & \text{Retail-Apparel/Shoe}
\\ 151 & \text{LTD} & \text{Ltd Brands Inc (LTD)} & \text{Consumer, Cyclical} & \text{Retail} & \text{Retail-Apparel/Shoe}
\\ 152 & \text{ROST} & \text{Ross Stores Inc (ROST)} & \text{Consumer, Cyclical} & \text{Retail} & \text{Retail-Apparel/Shoe}
\\ 153 & \text{URBN} & \text{Urban Outfitters Inc (URBN)} & \text{Consumer, Cyclical} & \text{Retail} & \text{Retail-Apparel/Shoe}
\\ 154 & \text{AZO} & \text{AutoZone Inc (AZO)} & \text{Consumer, Cyclical} & \text{Retail} & \text{Retail-Auto Parts}
\\ 155 & \text{ORLY} & \text{O'Reilly Automotive Inc (ORLY)} & \text{Consumer, Cyclical} & \text{Retail} & \text{Retail-Auto Parts}
\\ 156 & \text{AN} & \text{AutoNation Inc (AN)} & \text{Consumer, Cyclical} & \text{Retail} & \text{Retail-Automobile}
\\ 157 & \text{KMX} & \text{CarMax Inc (KMX)} & \text{Consumer, Cyclical} & \text{Retail} & \text{Retail-Automobile}
\\ 158 & \text{BBBY} & \text{Bed Bath \& Beyond Inc (BBBY)} & \text{Consumer, Cyclical} & \text{Retail} & \text{Retail-Bedding}
\\ 159 & \text{HD} & \text{Home Depot Inc/The (HD)} & \text{Consumer, Cyclical} & \text{Retail} & \text{Retail-Building Products}
\\ 160 & \text{LOW} & \text{Lowe's Cos Inc (LOW)} & \text{Consumer, Cyclical} & \text{Retail} & \text{Retail-Building Products}
\\ 161 & \text{GME} & \text{GameStop Corp (GME)} & \text{Consumer, Cyclical} & \text{Retail} & \text{Retail-Computer Equip}
\\ 162 & \text{BBY} & \text{Best Buy Co Inc (BBY)} & \text{Consumer, Cyclical} & \text{Retail} & \text{Retail-Consumer Electron}
\\ 163 & \text{BIG} & \text{Big Lots Inc (BIG)} & \text{Consumer, Cyclical} & \text{Retail} & \text{Retail-Discount}
\\ 164 & \text{DLTR} & \text{Dollar Tree Inc (DLTR)} & \text{Consumer, Cyclical} & \text{Retail} & \text{Retail-Discount}
\\ 165 & \text{FDO} & \text{Family Dollar Stores Inc (FDO)} & \text{Consumer, Cyclical} & \text{Retail} & \text{Retail-Discount}
\\ 166 & \text{TGT} & \text{Target Corp (TGT)} & \text{Consumer, Cyclical} & \text{Retail} & \text{Retail-Discount}
\\ 167 & \text{COST} & \text{Costco Wholesale Corp (COST)} & \text{Consumer, Cyclical} & \text{Retail} & \text{Retail-Discount}
\\ 168 & \text{WMT} & \text{Wal-Mart Stores Inc (WMT)} & \text{Consumer, Cyclical} & \text{Retail} & \text{Retail-Discount}
\\ 169 & \text{CVS} & \text{CVS Caremark Corp (CVS)} & \text{Consumer, Cyclical} & \text{Retail} & \text{Retail-Drug Store}
\\ 170 & \text{WAG} & \text{Walgreen Co (WAG)} & \text{Consumer, Cyclical} & \text{Retail} & \text{Retail-Drug Store}
\\ 171 & \text{TIF} & \text{Tiffany \& Co (TIF)} & \text{Consumer, Cyclical} & \text{Retail} & \text{Retail-Jewelry}
\\ 172 & \text{JWN} & \text{Nordstrom Inc (JWN)} & \text{Consumer, Cyclical} & \text{Retail} & \text{Retail-Major Dept Store}
\\ 173 & \text{JCP} & \text{JC Penney Co Inc (JCP)} & \text{Consumer, Cyclical} & \text{Retail} & \text{Retail-Major Dept Store}
\\ 174 & \text{TJX} & \text{TJX Cos Inc (TJX)} & \text{Consumer, Cyclical} & \text{Retail} & \text{Retail-Major Dept Store}
\\ 175 & \text{SPLS} & \text{Staples Inc (SPLS)} & \text{Consumer, Cyclical} & \text{Retail} & \text{Retail-Office Supplies}
\\ 176 & \text{KSS} & \text{Kohl's Corp (KSS)} & \text{Consumer, Cyclical} & \text{Retail} & \text{Retail-Regnl Dept Store}
\\ 177 & \text{M} & \text{Macy's Inc (M)} & \text{Consumer, Cyclical} & \text{Retail} & \text{Retail-Regnl Dept Store}
\\ 178 & \text{DRI} & \text{Darden Restaurants Inc (DRI)} & \text{Consumer, Cyclical} & \text{Retail} & \text{Retail-Restaurants}
\\ 179 & \text{MCD} & \text{McDonald's Corp (MCD)} & \text{Consumer, Cyclical} & \text{Retail} & \text{Retail-Restaurants}
\\ 180 & \text{SBUX} & \text{Starbucks Corp (SBUX)} & \text{Consumer, Cyclical} & \text{Retail} & \text{Retail-Restaurants}
\\ 181 & \text{YUM} & \text{Yum! Brands Inc (YUM)} & \text{Consumer, Cyclical} & \text{Retail} & \text{Retail-Restaurants}
\\ 182 & \text{CTAS} & \text{Cintas Corp (CTAS)} & \text{Consumer, Cyclical} & \text{Textyles} & \text{Linen Supply \& Rel Items}
\\ 183 & \text{HAS} & \text{Hasbro Inc (HAS)} & \text{Consumer, Cyclical} & \text{Toys / Games / Hobbies} & \text{Toys}
\\ 184 & \text{MAT} & \text{Mattel Inc (MAT)} & \text{Consumer, Cyclical} & \text{Toys / Games / Hobbies} & \text{Toys}
\\ 185 & \text{LUK} & \text{Leucadia National Corp (LUK)} & \text{Diversified} & \text{Holding Companies - Diversified} & \text{Diversified Operations}
\\ 186 & \text{TRV} & \text{Two Rivers Financial Group Inc (TRVR)} & \text{Financial} & \text{Banks} & \text{Commer Banks-Central US}
\\ 187 & \text{MTB} & \text{M \& T Bank Corp (MTB)} & \text{Financial} & \text{Banks} & \text{Commer Banks-Eastern US}
\\ 188 & \text{BBT} & \text{BB \& T Corp (BBT)} & \text{Financial} & \text{Banks} & \text{Commer Banks-Southern US}
\\ 189 & \text{FHN} & \text{First Horizon National Corp (FHN)} & \text{Financial} & \text{Banks} & \text{Commer Banks-Southern US}
\\ 190 & \text{RF} & \text{Regions Financial Corp (RF)} & \text{Financial} & \text{Banks} & \text{Commer Banks-Southern US}
\\ 191 & \text{ZION} & \text{Zions Bancorporation (ZION)} & \text{Financial} & \text{Banks} & \text{Commer Banks-Western US}
\\ 192 & \text{BAC} & \text{Bank of America Corp (BAC)} & \text{Financial} & \text{Banks} & \text{Diversified Banking Inst}
\\ 193 & \text{C} & \text{Citigroup Inc (C)} & \text{Financial} & \text{Banks} & \text{Diversified Banking Inst}
\\ 194 & \text{GS} & \text{Goldman Sachs Group Inc/The (GS)} & \text{Financial} & \text{Banks} & \text{Diversified Banking Inst}
\\ 195 & \text{JPM} & \text{JPMorgan Chase \& Co (JPM)} & \text{Financial} & \text{Banks} & \text{Diversified Banking Inst}
\end{array} \]

\[ \! \! \! \! \! \! \! \! \! \! \begin{array}{c|c|l|l|l|l} \\ & \text{Ticker} & \text{Company} & \text{Sector} & \text{Industry} & \text{Sub-Industry}
\\ \hline 196 & \text{MS} & \text{Morgan Stanley (MS)} & \text{Financial} & \text{Banks} & \text{Diversified Banking Inst}
\\ 197 & \text{BK} & \text{Bank of New York Mellon Corp/The (BK)} & \text{Financial} & \text{Banks} & \text{Fiduciary Banks}
\\ 198 & \text{NTRS} & \text{Northern Trust Corp (NTRS)} & \text{Financial} & \text{Banks} & \text{Fiduciary Banks}
\\ 199 & \text{STT} & \text{State Street Corp (STT)} & \text{Financial} & \text{Banks} & \text{Fiduciary Banks}
\\ 200 & \text{COF} & \text{Capital One Financial Corp (COF)} & \text{Financial} & \text{Banks} & \text{Super-Regional Banks-US}
\\ 201 & \text{CMA} & \text{Comerica Inc (CMA)} & \text{Financial} & \text{Banks} & \text{Super-Regional Banks-US}
\\ 202 & \text{FITB} & \text{Fifth Third Bancorp (FITB)} & \text{Financial} & \text{Banks} & \text{Super-Regional Banks-US}
\\ 203 & \text{HBAN} & \text{Huntington Bancshares Inc/OH (HBAN)} & \text{Financial} & \text{Banks} & \text{Super-Regional Banks-US}
\\ 204 & \text{KEY} & \text{KeyCorp (KEY)} & \text{Financial} & \text{Banks} & \text{Super-Regional Banks-US}
\\ 205 & \text{PNC} & \text{PNC Financial Services Group Inc (PNC)} & \text{Financial} & \text{Banks} & \text{Super-Regional Banks-US}
\\ 206 & \text{STI} & \text{SunTrust Banks Inc (STI)} & \text{Financial} & \text{Banks} & \text{Super-Regional Banks-US}
\\ 207 & \text{USB} & \text{US Bancorp (USB)} & \text{Financial} & \text{Banks} & \text{Super-Regional Banks-US}
\\ 208 & \text{WFC} & \text{Wells Fargo \& Co (WFC)} & \text{Financial} & \text{Banks} & \text{Super-Regional Banks-US}
\\ 209 & \text{SLM} & \text{SLM Corp (SLM)} & \text{Financial} & \text{Diversified Financial Services} & \text{Finance-Consumer Loans}
\\ 210 & \text{AXP} & \text{American Express Co (AXP)} & \text{Financial} & \text{Diversified Financial Services} & \text{Finance-Credit Card}
\\ 211 & \text{SCHW} & \text{Charles Schwab Corp/The (SCHW)} & \text{Financial} & \text{Diversified Financial Services} & \text{Finance-Invest Bnkr/Brkr}
\\ 212 & \text{ETFC} & \text{E*TRADE Financial Corp (ETFC)} & \text{Financial} & \text{Diversified Financial Services} & \text{Finance-Invest Bnkr/Brkr}
\\ 213 & \text{CME} & \text{CME Group Inc/IL (CME)} & \text{Financial} & \text{Diversified Financial Services} & \text{Finance-Other Services}
\\ 214 & \text{BLK } & \text{BlackRock Inc (BLK)} & \text{Financial} & \text{Diversified Financial Services} & \text{Invest Mgmnt/Advis Serv}
\\ 215 & \text{FII} & \text{Federated Investors Inc (FII)} & \text{Financial} & \text{Diversified Financial Services} & \text{Invest Mgmnt/Advis Serv}
\\ 216 & \text{BEN} & \text{Franklin Resources Inc (BEN)} & \text{Financial} & \text{Diversified Financial Services} & \text{Invest Mgmnt/Advis Serv}
\\ 217 & \text{IVZ} & \text{Invesco Ltd (IVZ)} & \text{Financial} & \text{Diversified Financial Services} & \text{Invest Mgmnt/Advis Serv}
\\ 218 & \text{LM} & \text{Legg Mason Inc (LM)} & \text{Financial} & \text{Diversified Financial Services} & \text{Invest Mgmnt/Advis Serv}
\\ 219 & \text{TROW} & \text{T Rowe Price Group Inc (TROW)} & \text{Financial} & \text{Diversified Financial Services} & \text{Invest Mgmnt/Advis Serv}
\\ 220 & \text{AON} & \text{Aon PLC (AON)} & \text{Financial} & \text{Insurance} & \text{Insurance Brokers}
\\ 221 & \text{MMC} & \text{Marsh \& McLennan Cos Inc (MMC)} & \text{Financial} & \text{Insurance} & \text{Insurance Brokers}
\\ 222 & \text{AFL} & \text{Aflac Inc (AFL)} & \text{Financial} & \text{Insurance} & \text{Life/Health Insurance}
\\ 223 & \text{LNC} & \text{Lincoln National Corp (LNC)} & \text{Financial} & \text{Insurance} & \text{Life/Health Insurance}
\\ 224 & \text{PFG} & \text{Principal Financial Group Inc (PFG)} & \text{Financial} & \text{Insurance} & \text{Life/Health Insurance}
\\ 225 & \text{PRU} & \text{Prudential Financial Inc (PRU)} & \text{Financial} & \text{Insurance} & \text{Life/Health Insurance}
\\ 226 & \text{TMK} & \text{Torchmark Corp (TMK)} & \text{Financial} & \text{Insurance} & \text{Life/Health Insurance}
\\ 227 & \text{UNM} & \text{Unum Group (UNM)} & \text{Financial} & \text{Insurance} & \text{Life/Health Insurance}
\\ 228 & \text{ACE} & \text{ACE Ltd (ACE)} & \text{Financial} & \text{Insurance} & \text{Multi-line Insurance}
\\ 229 & \text{ALL} & \text{Allstate Corp/The (ALL)} & \text{Financial} & \text{Insurance} & \text{Multi-line Insurance}
\\ 230 & \text{AIG} & \text{American International Group Inc (AIG)} & \text{Financial} & \text{Insurance} & \text{Multi-line Insurance}
\\ 231 & \text{CINF} & \text{Cincinnati Financial Corp (CINF)} & \text{Financial} & \text{Insurance} & \text{Multi-line Insurance}
\\ 232 & \text{HIG} & \text{Hartford Financial Services Group Inc (HIG)} & \text{Financial} & \text{Insurance} & \text{Multi-line Insurance}
\\ 233 & \text{L} & \text{Loews Corp (L)} & \text{Financial} & \text{Insurance} & \text{Multi-line Insurance}
\\ 234 & \text{MET} & \text{MetLife Inc (MET)} & \text{Financial} & \text{Insurance} & \text{Multi-line Insurance}
\\ 235 & \text{XL} & \text{XL Group PLC (XL)} & \text{Financial} & \text{Insurance} & \text{Multi-line Insurance}
\\ 236 & \text{CB} & \text{Chubb Corp/The (CB)} & \text{Financial} & \text{Insurance} & \text{Property/Casualty Ins}
\\ 237 & \text{PGR} & \text{Progressive Corp/The (PGR)} & \text{Financial} & \text{Insurance} & \text{Property/Casualty Ins}
\\ 238 & \text{BRK/B} & \text{Berkshire Hathaway Inc (BRK/B)} & \text{Financial} & \text{Insurance} & \text{Reinsurance}
\\ 239 & \text{AIV} & \text{Apartment Investment \& Management Co (AIV)} & \text{Financial} & \text{REITS} & \text{REITS-Apartments}
\\ 240 & \text{AVB} & \text{AvalonBay Communities Inc (AVB)} & \text{Financial} & \text{REITS} & \text{REITS-Apartments}
\\ 241 & \text{EQR} & \text{Equity Residential (EQR)} & \text{Financial} & \text{REITS} & \text{REITS-Apartments}
\\ 242 & \text{PCL} & \text{Plum Creek Timber Co Inc (PCL)} & \text{Financial} & \text{REITS} & \text{REITS-Diversified}
\\ 243 & \text{VNO} & \text{Vornado Realty Trust (VNO)} & \text{Financial} & \text{REITS} & \text{REITS-Diversified}
\\ 244 & \text{WY} & \text{Weyerhaeuser Co (WY)} & \text{Financial} & \text{REITS} & \text{REITS-Diversified}
\\ 245 & \text{AMT} & \text{American Tower Corp (AMT)} & \text{Financial} & \text{REITS} & \text{REITS-Diversified}
\\ 246 & \text{HCP} & \text{HCP Inc (HCP)} & \text{Financial} & \text{REITS} & \text{REITS-Health Care}
\\ 247 & \text{HCN} & \text{Health Care REIT Inc (HCN)} & \text{Financial} & \text{REITS} & \text{REITS-Health Care}
\\ 248 & \text{VTR} & \text{Ventas Inc (VTR)} & \text{Financial} & \text{REITS} & \text{REITS-Health Care}
\\ 249 & \text{HST} & \text{Host Hotels \& Resorts Inc (HST)} & \text{Financial} & \text{REITS} & \text{REITS-Hotels}
\\ 250 & \text{BXP} & \text{Boston Properties Inc (BXP)} & \text{Financial} & \text{REITS} & \text{REITS-Office Property}
\\ 251 & \text{SPG} & \text{Simon Property Group Inc (SPG)} & \text{Financial} & \text{REITS} & \text{REITS-Regional Malls}
\\ 252 & \text{KIM} & \text{Kimco Realty Corp (KIM)} & \text{Financial} & \text{REITS} & \text{REITS-Shopping Centers}
\\ 253 & \text{PSA} & \text{Public Storage (PSA)} & \text{Financial} & \text{REITS} & \text{REITS-Storage}
\\ 254 & \text{PLD} & \text{Prologis Inc (PLD)} & \text{Financial} & \text{REITS} & \text{REITS-Warehouse/Industr}
\\ 255 & \text{HCBK} & \text{Hudson City Bancorp Inc (HCBK)} & \text{Financial} & \text{Savings \& Loans} & \text{S \& L/Thrifts-Eastern US}
\\ 256 & \text{PBCT} & \text{People's United Financial Inc (PBCT)} & \text{Financial} & \text{Savings \& Loans} & \text{S \& L/Thrifts-Eastern US}
\\ 257 & \text{IPG} & \text{Interpublic Group of Cos Inc/The (IPG)} & \text{Communications} & \text{Advertising} & \text{Advertising Agencies}
\\ 258 & \text{OMC} & \text{Omnicom Group Inc (OMC)} & \text{Communications} & \text{Advertising} & \text{Advertising Agencies}
\\ 259 & \text{AMZN} & \text{Amazon.com Inc (AMZN)} & \text{Communications} & \text{Internet} & \text{E-Commerce/Products}
\\ 260 & \text{EBAY} & \text{eBay Inc (EBAY)} & \text{Communications} & \text{Internet} & \text{E-Commerce/Products}
\\ 261 & \text{NFLX} & \text{Netflix Inc (NFLX)} & \text{Communications} & \text{Internet} & \text{E-Commerce/Products}
\\ 262 & \text{PCLN} & \text{priceline.com Inc (PCLN)} & \text{Communications} & \text{Internet} & \text{E-Commerce/Services}
\\ 263 & \text{FFIV} & \text{F5 Networks Inc (FFIV)} & \text{Communications} & \text{Internet} & \text{Internet Infrastr Sftwr}
\\ 264 & \text{SYMC} & \text{Symantec Corp (SYMC)} & \text{Communications} & \text{Internet} & \text{Internet Security}
\\ 265 & \text{VRSN} & \text{VeriSign Inc (VRSN)} & \text{Communications} & \text{Internet} & \text{Internet Security}
\\ 266 & \text{YHOO} & \text{Yahoo! Inc (YHOO)} & \text{Communications} & \text{Internet} & \text{Web Portals/ISP}
\\ 267 & \text{CVC} & \text{Cablevision Systems Corp (CVC)} & \text{Communications} & \text{Media} & \text{Cable/Satellite TV}
\\ 268 & \text{CMCSA} & \text{Comcast Corp (CMCSA)} & \text{Communications} & \text{Media} & \text{Cable/Satellite TV}
\\ 269 & \text{DTV} & \text{DIRECTV (DTV)} & \text{Communications} & \text{Media} & \text{Cable/Satellite TV}
\\ 270 & \text{NWSA} & \text{News Corp (NWSA)} & \text{Communications} & \text{Media} & \text{Multimedia}
\\ 271 & \text{TWX} & \text{Time Warner Inc (TWX)} & \text{Communications} & \text{Media} & \text{Multimedia}
\\ 272 & \text{DIS} & \text{Walt Disney Co/The (DIS)} & \text{Communications} & \text{Media} & \text{Multimedia}
\\ 273 & \text{MHP} & \text{McGraw-Hill Cos Inc/The (MHP)} & \text{Communications} & \text{Media} & \text{Publishing-Books}
\\ 274 & \text{GCI} & \text{Gannett Co Inc (GCI)} & \text{Communications} & \text{Media} & \text{Publishing-Newspapers}
\\ 275 & \text{WPO} & \text{Washington Post Co/The (WPO)} & \text{Communications} & \text{Media} & \text{Publishing-Newspapers}
\\ 276 & \text{SIRI} & \text{Sirius XM Radio Inc (SIRI)} & \text{Communications} & \text{Media} & \text{Radio}
\\ 277 & \text{CBS} & \text{CBS Corp (CBS)} & \text{Communications} & \text{Media} & \text{Television}
\\ 278 & \text{S} & \text{Sprint Nextel Corp (S)} & \text{Communications} & \text{Telecommunications} & \text{Cellular Telecom}
\\ 279 & \text{VOD} & \text{Vodafone Group PLC (VOD)} & \text{Communications} & \text{Telecommunications} & \text{Cellular Telecom}
\\ 280 & \text{CSCO} & \text{Cisco Systems Inc (CSCO)} & \text{Communications} & \text{Telecommunications} & \text{Networking Products}
\\ 281 & \text{GLW} & \text{Corning Inc (GLW)} & \text{Communications} & \text{Telecommunications} & \text{Telecom Eq Fiber Optics}
\\ 282 & \text{JDSU} & \text{JDS Uniphase Corp (JDSU)} & \text{Communications} & \text{Telecommunications} & \text{Telecom Eq Fiber Optics}
\\ 283 & \text{HRS} & \text{Harris Corp (HRS)} & \text{Communications} & \text{Telecommunications} & \text{Telecommunication Equip}
\\ 284 & \text{JNPR} & \text{Juniper Networks Inc (JNPR)} & \text{Communications} & \text{Telecommunications} & \text{Telecommunication Equip}
\\ 285 & \text{T} & \text{AT \& T Inc (T)} & \text{Communications} & \text{Telecommunications} & \text{Telephone-Integrated}
\\ 286 & \text{CTL} & \text{CenturyLink Inc (CTL)} & \text{Communications} & \text{Telecommunications} & \text{Telephone-Integrated}
\\ 287 & \text{FTR} & \text{Frontier Communications Corp (FTR)} & \text{Communications} & \text{Telecommunications} & \text{Telephone-Integrated}
\\ 288 & \text{VZ} & \text{Verizon Communications Inc (VZ)} & \text{Communications} & \text{Telecommunications} & \text{Telephone-Integrated}
\\ 289 & \text{MSI} & \text{Motorola Solutions Inc (MSI)} & \text{Communications} & \text{Telecommunications} & \text{Wireless Equipment}
\\ 290 & \text{SBAC} & \text{SBA Communications Corp (SBAC)} & \text{Communications} & \text{Telecommunications} & \text{Wireless Equipment}
\\ 291 & \text{ACN} & \text{Accenture PLC (ACN)} & \text{Technology} & \text{Computers} & \text{Computer Services}
\\ 292 & \text{CTSH} & \text{Cognizant Technology Solutions Corp (CTSH)} & \text{Technology} & \text{Computers} & \text{Computer Services}
\\ 293 & \text{CSC} & \text{Computer Sciences Corp (CSC)} & \text{Technology} & \text{Computers} & \text{Computer Services}
\end{array} \]

\[ \! \! \! \! \! \! \! \! \! \! \begin{array}{c|c|l|l|l|l} \\ & \text{Ticker} & \text{Company} & \text{Sector} & \text{Industry} & \text{Sub-Industry}
\\ \hline 294 & \text{IBM} & \text{International Business Machines Corp (IBM)} & \text{Technology} & \text{Computers} & \text{Computer Services}
\\ 295 & \text{AAPL} & \text{Apple Inc (AAPL)} & \text{Technology} & \text{Computers} & \text{Computers}
\\ 296 & \text{DELL} & \text{Dell Inc (DELL)} & \text{Technology} & \text{Computers} & \text{Computers}
\\ 297 & \text{HPQ} & \text{Hewlett-Packard Co (HPQ)} & \text{Technology} & \text{Computers} & \text{Computers}
\\ 298 & \text{EMC} & \text{EMC Corp/MA (EMC)} & \text{Technology} & \text{Computers} & \text{Computers-Memory Devices}
\\ 299 & \text{NTAP} & \text{NetApp Inc (NTAP)} & \text{Technology} & \text{Computers} & \text{Computers-Memory Devices}
\\ 300 & \text{SNDK} & \text{SanDisk Corp (SNDK)} & \text{Technology} & \text{Computers} & \text{Computers-Memory Devices}
\\ 301 & \text{WDC} & \text{Western Digital Corp (WDC)} & \text{Technology} & \text{Computers} & \text{Computers-Memory Devices}
\\ 302 & \text{STX} & \text{Seagate Technology PLC (STX)} & \text{Technology} & \text{Computers} & \text{Computers-Memory Devices}
\\ 303 & \text{LXK} & \text{Lexmark International Inc (LXK)} & \text{Technology} & \text{Computers} & \text{Computers-Peripher Equip}
\\ 304 & \text{PBI} & \text{Pitney Bowes Inc (PBI)} & \text{Technology} & \text{Office / Business Equipment} & \text{Office Automation \& Equip}
\\ 305 & \text{XRX} & \text{Xerox Corp (XRX)} & \text{Technology} & \text{Office / Business Equipment} & \text{Office Automation \& Equip}
\\ 306 & \text{AMD} & \text{Advanced Micro Devices Inc (AMD)} & \text{Technology} & \text{Semiconductors} & \text{Electronic Compo-Semicon}
\\ 307 & \text{ALTR} & \text{Altera Corp (ALTR)} & \text{Technology} & \text{Semiconductors} & \text{Electronic Compo-Semicon}
\\ 308 & \text{BRCM} & \text{Broadcom Corp (BRCM)} & \text{Technology} & \text{Semiconductors} & \text{Electronic Compo-Semicon}
\\ 309 & \text{INTC} & \text{Intel Corp (INTC)} & \text{Technology} & \text{Semiconductors} & \text{Electronic Compo-Semicon}
\\ 310 & \text{LSI} & \text{LSI Corp (LSI)} & \text{Technology} & \text{Semiconductors} & \text{Electronic Compo-Semicon}
\\ 311 & \text{MCHP} & \text{Microchip Technology Inc (MCHP)} & \text{Technology} & \text{Semiconductors} & \text{Electronic Compo-Semicon}
\\ 312 & \text{MU} & \text{Micron Technology Inc (MU)} & \text{Technology} & \text{Semiconductors} & \text{Electronic Compo-Semicon}
\\ 313 & \text{NVDA} & \text{NVIDIA Corp (NVDA)} & \text{Technology} & \text{Semiconductors} & \text{Electronic Compo-Semicon}
\\ 314 & \text{TXN} & \text{Texas Instruments Inc (TXN)} & \text{Technology} & \text{Semiconductors} & \text{Electronic Compo-Semicon}
\\ 315 & \text{XLNX} & \text{Xilinx Inc (XLNX)} & \text{Technology} & \text{Semiconductors} & \text{Electronic Compo-Semicon}
\\ 316 & \text{ADI} & \text{Analog Devices Inc (ADI)} & \text{Technology} & \text{Semiconductors} & \text{Semicon Compo-Intg Circu}
\\ 317 & \text{LLTC} & \text{Linear Technology Corp (LLTC)} & \text{Technology} & \text{Semiconductors} & \text{Semicon Compo-Intg Circu}
\\ 318 & \text{QCOM} & \text{QUALCOMM Inc (QCOM)} & \text{Technology} & \text{Semiconductors} & \text{Semicon Compo-Intg Circu}
\\ 319 & \text{MXIM} & \text{Maxim Integrated Products Inc (MXIM)} & \text{Technology} & \text{Semiconductors} & \text{Semicon Compo-Intg Circu}
\\ 320 & \text{AMAT} & \text{Applied Materials Inc (AMAT)} & \text{Technology} & \text{Semiconductors} & \text{Semiconductor Equipment}
\\ 321 & \text{KLAC} & \text{KLA-Tencor Corp (KLAC)} & \text{Technology} & \text{Semiconductors} & \text{Semiconductor Equipment}
\\ 322 & \text{TER} & \text{Teradyne Inc (TER)} & \text{Technology} & \text{Semiconductors} & \text{Semiconductor Equipment}
\\ 323 & \text{CTXS} & \text{Citrix Systems Inc (CTXS)} & \text{Technology} & \text{Software} & \text{Applications Software}
\\ 324 & \text{CPWR} & \text{Compuware Corp (CPWR)} & \text{Technology} & \text{Software} & \text{Applications Software}
\\ 325 & \text{INTU} & \text{Intuit Inc (INTU)} & \text{Technology} & \text{Software} & \text{Applications Software}
\\ 326 & \text{MSFT} & \text{Microsoft Corp (MSFT)} & \text{Technology} & \text{Software} & \text{Applications Software}
\\ 327 & \text{RHT} & \text{Red Hat Inc (RHT)} & \text{Technology} & \text{Software} & \text{Applications Software}
\\ 328 & \text{CHKP} & \text{Check Point Software Technologies Ltd (CHKP)} & \text{Technology} & \text{Software} & \text{Applications Software}
\\ 329 & \text{NUAN} & \text{Nuance Communications Inc (NUAN)} & \text{Technology} & \text{Software} & \text{Applications Software}
\\ 330 & \text{ADSK} & \text{Autodesk Inc (ADSK)} & \text{Technology} & \text{Software} & \text{Computer Aided Design}
\\ 331 & \text{AKAM} & \text{Akamai Technologies Inc (AKAM)} & \text{Technology} & \text{Software} & \text{Computer Software}
\\ 332 & \text{DNB} & \text{Dun \& Bradstreet Corp/The (DNB)} & \text{Technology} & \text{Software} & \text{Data Processing/Mgmt}
\\ 333 & \text{FIS} & \text{Fidelity National Information Services Inc (FIS)} & \text{Technology} & \text{Software} & \text{Data Processing/Mgmt}
\\ 334 & \text{FISV} & \text{Fiserv Inc (FISV)} & \text{Technology} & \text{Software} & \text{Data Processing/Mgmt}
\\ 335 & \text{ADBE} & \text{Adobe Systems Inc (ADBE)} & \text{Technology} & \text{Software} & \text{Electronic Forms}
\\ 336 & \text{BMC} & \text{BMC Software Inc (BMC)} & \text{Technology} & \text{Software} & \text{Enterprise Software/Serv}
\\ 337 & \text{CA} & \text{CA Inc (CA)} & \text{Technology} & \text{Software} & \text{Enterprise Software/Serv}
\\ 338 & \text{ORCL} & \text{Oracle Corp (ORCL)} & \text{Technology} & \text{Software} & \text{Enterprise Software/Serv}
\\ 339 & \text{EA} & \text{Electronic Arts Inc (EA)} & \text{Technology} & \text{Software} & \text{Entertainment Software}
\\ 340 & \text{ATVI} & \text{Activision Blizzard Inc (ATVI)} & \text{Technology} & \text{Software} & \text{Entertainment Software}
\\ 341 & \text{CERN} & \text{Cerner Corp (CERN)} & \text{Technology} & \text{Software} & \text{Medical Information Sys}
\\ 342 & \text{AES} & \text{AES Corp/VA (AES)} & \text{Utilities} & \text{Electric} & \text{Electric-Generation}
\\ 343 & \text{AEE} & \text{Ameren Corp (AEE)} & \text{Utilities} & \text{Electric} & \text{Electric-Integrated}
\\ 344 & \text{AEP} & \text{American Electric Power Co Inc (AEP)} & \text{Utilities} & \text{Electric} & \text{Electric-Integrated}
\\ 345 & \text{CMS} & \text{CMS Energy Corp (CMS)} & \text{Utilities} & \text{Electric} & \text{Electric-Integrated}
\\ 346 & \text{ED} & \text{Consolidated Edison Inc (ED)} & \text{Utilities} & \text{Electric} & \text{Electric-Integrated}
\\ 347 & \text{D} & \text{Dominion Resources Inc/VA (D)} & \text{Utilities} & \text{Electric} & \text{Electric-Integrated}
\\ 348 & \text{DTE} & \text{DTE Energy Co (DTE)} & \text{Utilities} & \text{Electric} & \text{Electric-Integrated}
\\ 349 & \text{DUK} & \text{Duke Energy Corp (DUK)} & \text{Utilities} & \text{Electric} & \text{Electric-Integrated}
\\ 350 & \text{EIX} & \text{Edison International (EIX)} & \text{Utilities} & \text{Electric} & \text{Electric-Integrated}
\\ 351 & \text{ETR} & \text{Entergy Corp (ETR)} & \text{Utilities} & \text{Electric} & \text{Electric-Integrated}
\\ 352 & \text{EXC} & \text{Exelon Corp (EXC)} & \text{Utilities} & \text{Electric} & \text{Electric-Integrated}
\\ 353 & \text{FE} & \text{FirstEnergy Corp (FE)} & \text{Utilities} & \text{Electric} & \text{Electric-Integrated}
\\ 354 & \text{TEG} & \text{Integrys Energy Group Inc (TEG)} & \text{Utilities} & \text{Electric} & \text{Electric-Integrated}
\\ 355 & \text{NEE} & \text{NextEra Energy Inc (NEE)} & \text{Utilities} & \text{Electric} & \text{Electric-Integrated}
\\ 356 & \text{NU} & \text{Northeast Utilities (NU)} & \text{Utilities} & \text{Electric} & \text{Electric-Integrated}
\\ 357 & \text{POM} & \text{Pepco Holdings Inc (POM)} & \text{Utilities} & \text{Electric} & \text{Electric-Integrated}
\\ 358 & \text{PCG} & \text{PG \& E Corp (PCG)} & \text{Utilities} & \text{Electric} & \text{Electric-Integrated}
\\ 359 & \text{PNW} & \text{Pinnacle West Capital Corp (PNW)} & \text{Utilities} & \text{Electric} & \text{Electric-Integrated}
\\ 360 & \text{PPL} & \text{PPL Corp (PPL)} & \text{Utilities} & \text{Electric} & \text{Electric-Integrated}
\\ 361 & \text{PEG} & \text{Public Service Enterprise Group Inc (PEG)} & \text{Utilities} & \text{Electric} & \text{Electric-Integrated}
\\ 362 & \text{SCG} & \text{SCANA Corp (SCG)} & \text{Utilities} & \text{Electric} & \text{Electric-Integrated}
\\ 363 & \text{SO} & \text{Southern Co/The (SO)} & \text{Utilities} & \text{Electric} & \text{Electric-Integrated}
\\ 364 & \text{TE} & \text{TECO Energy Inc (TE)} & \text{Utilities} & \text{Electric} & \text{Electric-Integrated}
\\ 365 & \text{WEC} & \text{Wisconsin Energy Corp (WEC)} & \text{Utilities} & \text{Electric} & \text{Electric-Integrated}
\\ 366 & \text{XEL} & \text{Xcel Energy Inc (XEL)} & \text{Utilities} & \text{Electric} & \text{Electric-Integrated}
\\ 367 & \text{GAS} & \text{AGL Resources Inc (GAS)} & \text{Utilities} & \text{Gas} & \text{Gas-Distribution}
\\ 368 & \text{CNP} & \text{CenterPoint Energy Inc (CNP)} & \text{Utilities} & \text{Gas} & \text{Gas-Distribution}
\\ 369 & \text{NI} & \text{NiSource Inc (NI)} & \text{Utilities} & \text{Gas} & \text{Gas-Distribution}
\\ 370 & \text{SRE} & \text{Sempra Energy (SRE)} & \text{Utilities} & \text{Gas} & \text{Gas-Distribution}
\\ 371 & \text{ADM} & \text{Archer-Daniels-Midland Co (ADM)} & \text{Consumer, Non-Cyclical} & \text{Agriculture} & \text{Agricultural Operations}
\\ 372 & \text{MO} & \text{Altria Group Inc (MO)} & \text{Consumer, Non-Cyclical} & \text{Agriculture} & \text{Tobacco}
\\ 373 & \text{RAI} & \text{Reynolds American Inc (RAI)} & \text{Consumer, Non-Cyclical} & \text{Agriculture} & \text{Tobacco}
\\ 374 & \text{KO} & \text{Coca-Cola Co/The (KO)} & \text{Consumer, Non-Cyclical} & \text{Beverages} & \text{Beverages-Non-alcoholic}
\\ 375 & \text{CCE} & \text{Coca-Cola Enterprises Inc (CCE)} & \text{Consumer, Non-Cyclical} & \text{Beverages} & \text{Beverages-Non-alcoholic}
\\ 376 & \text{PEP} & \text{PepsiCo Inc (PEP)} & \text{Consumer, Non-Cyclical} & \text{Beverages} & \text{Beverages-Non-alcoholic}
\\ 377 & \text{MNST} & \text{Monster Beverage Corp (MNST)} & \text{Consumer, Non-Cyclical} & \text{Beverages} & \text{Beverages-Non-alcoholic}
\\ 378 & \text{BEAM} & \text{Beam Inc (BEAM)} & \text{Consumer, Non-Cyclical} & \text{Beverages} & \text{Beverages-Wine/Spirits}
\\ 379 & \text{BF/B} & \text{Brown-Forman Corp (BF/B)} & \text{Consumer, Non-Cyclical} & \text{Beverages} & \text{Beverages-Wine/Spirits}
\\ 380 & \text{STZ} & \text{Constellation Brands Inc (STZ)} & \text{Consumer, Non-Cyclical} & \text{Beverages} & \text{Beverages-Wine/Spirits}
\\ 381 & \text{TAP} & \text{Molson Coors Brewing Co (TAP)} & \text{Consumer, Non-Cyclical} & \text{Beverages} & \text{Brewery}
\\ 382 & \text{AMGN} & \text{Amgen Inc (AMGN)} & \text{Consumer, Non-Cyclical} & \text{Biotechnology} & \text{Medical-Biomedical/Gene}
\\ 383 & \text{BIIB} & \text{Biogen Idec Inc (BIIB)} & \text{Consumer, Non-Cyclical} & \text{Biotechnology} & \text{Medical-Biomedical/Gene}
\\ 384 & \text{CELG} & \text{Celgene Corp (CELG)} & \text{Consumer, Non-Cyclical} & \text{Biotechnology} & \text{Medical-Biomedical/Gene}
\\ 385 & \text{GILD} & \text{Gilead Sciences Inc (GILD)} & \text{Consumer, Non-Cyclical} & \text{Biotechnology} & \text{Medical-Biomedical/Gene}
\\ 386 & \text{LIFE} & \text{Life Technologies Corp (LIFE)} & \text{Consumer, Non-Cyclical} & \text{Biotechnology} & \text{Medical-Biomedical/Gene}
\\ 387 & \text{ALXN} & \text{Alexion Pharmaceuticals Inc (ALXN)} & \text{Consumer, Non-Cyclical} & \text{Biotechnology} & \text{Medical-Biomedical/Gene}
\\ 388 & \text{REGN} & \text{Regeneron Pharmaceuticals Inc (REGN)} & \text{Consumer, Non-Cyclical} & \text{Biotechnology} & \text{Medical-Biomedical/Gene}
\\ 389 & \text{VRTX} & \text{Vertex Pharmaceuticals Inc (VRTX)} & \text{Consumer, Non-Cyclical} & \text{Biotechnology} & \text{Medical-Biomedical/Gene}
\\ 390 & \text{HRB} & \text{H \& R Block Inc (HRB)} & \text{Consumer, Non-Cyclical} & \text{Commercial Services} & \text{Commercial Serv-Finance}
\\ 391 & \text{EFX} & \text{Equifax Inc (EFX)} & \text{Consumer, Non-Cyclical} & \text{Commercial Services} & \text{Commercial Serv-Finance}
\end{array} \]

\[ \! \! \! \! \! \! \! \! \! \! \begin{array}{c|c|l|l|l|l} \\ & \text{Ticker} & \text{Company} & \text{Sector} & \text{Industry} & \text{Sub-Industry}
\\ \hline 392 & \text{MCO} & \text{Moody's Corp (MCO)} & \text{Consumer, Non-Cyclical} & \text{Commercial Services} & \text{Commercial Serv-Finance}
\\ 393 & \text{ADP} & \text{Automatic Data Processing Inc (ADP)} & \text{Consumer, Non-Cyclical} & \text{Commercial Services} & \text{Commercial Serv-Finance}
\\ 394 & \text{PAYX} & \text{Paychex Inc (PAYX)} & \text{Consumer, Non-Cyclical} & \text{Commercial Services} & \text{Commercial Serv-Finance}
\\ 395 & \text{TSS} & \text{Total System Services Inc (TSS)} & \text{Consumer, Non-Cyclical} & \text{Commercial Services} & \text{Commercial Serv-Finance}
\\ 396 & \text{IRM} & \text{Iron Mountain Inc (IRM)} & \text{Consumer, Non-Cyclical} & \text{Commercial Services} & \text{Commercial Services}
\\ 397 & \text{PWR} & \text{Quanta Services Inc (PWR)} & \text{Consumer, Non-Cyclical} & \text{Commercial Services} & \text{Commercial Services}
\\ 398 & \text{RHI} & \text{Robert Half International Inc (RHI)} & \text{Consumer, Non-Cyclical} & \text{Commercial Services} & \text{Human Resources}
\\ 399 & \text{RRD} & \text{RR Donnelley \& Sons Co (RRD)} & \text{Consumer, Non-Cyclical} & \text{Commercial Services} & \text{Printing-Commercial}
\\ 400 & \text{APOL} & \text{Apollo Group Inc (APOL)} & \text{Consumer, Non-Cyclical} & \text{Commercial Services} & \text{Schools}
\\ 401 & \text{DV} & \text{DeVry Inc (DV)} \& \text{Consumer, Non-Cyclical} & \text{Commercial Services} & \text{Schools}
\\ 402 & \text{AVP} & \text{Avon Products Inc (AVP)} & \text{Consumer, Non-Cyclical} & \text{Cosmetics / Personal Care} & \text{Cosmetics \& Toiletries}
\\ 403 & \text{CL} & \text{Colgate-Palmolive Co (CL)} & \text{Consumer, Non-Cyclical} & \text{Cosmetics / Personal Care} & \text{Cosmetics \& Toiletries}
\\ 404 & \text{EL} & \text{Estee Lauder Cos Inc/The (EL)} & \text{Consumer, Non-Cyclical} & \text{Cosmetics / Personal Care} & \text{Cosmetics \& Toiletries}
\\ 405 & \text{PG} & \text{Procter \& Gamble Co/The (PG)} & \text{Consumer, Non-Cyclical} & \text{Cosmetics / Personal Care} & \text{Cosmetics \& Toiletries}
\\ 406 & \text{SJM} & \text{JM Smucker Co/The (SJM)} & \text{Consumer, Non-Cyclical} & \text{Food} & \text{Food-Confectionery}
\\ 407 & \text{HSY} & \text{Hershey Co/The (HSY)} & \text{Consumer, Non-Cyclical} & \text{Food} & \text{Food-Confectionery}
\\ 408 & \text{DF} & \text{Dean Foods Co (DF)} & \text{Consumer, Non-Cyclical} & \text{Food} & \text{Food-Dairy Products}
\\ 409 & \text{HRL} & \text{Hormel Foods Corp (HRL)} & \text{Consumer, Non-Cyclical} & \text{Food} & \text{Food-Meat Products}
\\ 410 & \text{HSH} & \text{Hillshire Brands Co (HSH)} & \text{Consumer, Non-Cyclical} & \text{Food} & \text{Food-Meat Products}
\\ 411 & \text{TSN} & \text{Tyson Foods Inc (TSN)} & \text{Consumer, Non-Cyclical} & \text{Food} & \text{Food-Meat Products}
\\ 412 & \text{ CPB} & \text{Campbell Soup Co (CPB)} & \text{Consumer, Non-Cyclical} & \text{Food} & \text{Food-Misc/Diversified}
\\ 413 & \text{CAG} & \text{ConAgra Foods Inc (CAG)} & \text{Consumer, Non-Cyclical} & \text{Food} & \text{Food-Misc/Diversified}
\\ 414 & \text{GIS} & \text{General Mills Inc (GIS)} & \text{Consumer, Non-Cyclical} & \text{Food} & \text{Food-Misc/Diversified}
\\ 415 & \text{HNZ} & \text{HJ Heinz Co (HNZ)} & \text{Consumer, Non-Cyclical} & \text{Food} & \text{Food-Misc/Diversified}
\\ 416 & \text{K} & \text{Kellogg Co (K)} & \text{Consumer, Non-Cyclical} & \text{Food} & \text{Food-Misc/Diversified}
\\ 417 & \text{MKC} & \text{McCormick \& Co Inc/MD (MKC)} & \text{Consumer, Non-Cyclical} & \text{Food} & \text{Food-Misc/Diversified}
\\ 418 & \text{KR} & \text{Kroger Co/The (KR)} & \text{Consumer, Non-Cyclical} & \text{Food} & \text{Food-Retail}
\\ 419 & \text{SWY} & \text{Safeway Inc (SWY)} & \text{Consumer, Non-Cyclical} & \text{Food} & \text{Food-Retail}
\\ 420 & \text{SVU} & \text{SUPERVALU Inc (SVU)} & \text{Consumer, Non-Cyclical} & \text{Food} & \text{Food-Retail}
\\ 421 & \text{WFM} & \text{Whole Foods Market Inc (WFM)} & \text{Consumer, Non-Cyclical} & \text{Food} & \text{Food-Retail}
\\ 422 & \text{SYY} & \text{Sysco Corp (SYY)} & \text{Consumer, Non-Cyclical} & \text{Food} & \text{Food-Wholesale/Distrib}
\\ 423 & \text{XRAY} & \text{DENTSPLY International Inc (XRAY)} & \text{Consumer, Non-Cyclical} & \text{Health Care - Products} & \text{Dental Supplies \& Equip}
\\ 424 & \text{PDCO} & \text{Patterson Cos Inc (PDCO)} & \text{Consumer, Non-Cyclical} & \text{Health Care - Products} & \text{Dental Supplies \& Equip}
\\ 425 & \text{BCR} & \text{CR Bard Inc (BCR)} & \text{Consumer, Non-Cyclical} & \text{Health Care - Products} & \text{Disposable Medical Prod}
\\ 426 & \text{BSX} & \text{Boston Scientific Corp (BSX)} & \text{Consumer, Non-Cyclical} & \text{Health Care - Products} & \text{Medical Instruments}
\\ 427 & \text{EW} & \text{Edwards Lifesciences Corp (EW)} & \text{Consumer, Non-Cyclical} & \text{Health Care - Products} & \text{Medical Instruments}
\\ 428 & \text{ISRG} & \text{Intuitive Surgical Inc (ISRG)} & \text{Consumer, Non-Cyclical} & \text{Health Care - Products} & \text{Medical Instruments}
\\ 429 & \text{MDT} & \text{Medtronic Inc (MDT)} & \text{Consumer, Non-Cyclical} & \text{Health Care - Products} & \text{Medical Instruments}
\\ 430 & \text{STJ} & \text{St Jude Medical Inc (STJ)} & \text{Consumer, Non-Cyclical} & \text{Health Care - Products} & \text{Medical Instruments}
\\ 431 & \text{BAX} & \text{Baxter International Inc (BAX)} & \text{Consumer, Non-Cyclical} & \text{Health Care - Products} & \text{Medical Products}
\\ 432 & \text{BDX} & \text{Becton Dickinson and Co (BDX)} & \text{Consumer, Non-Cyclical} & \text{Health Care - Products} & \text{Medical Products}
\\ 433 & \text{SYK} & \text{Stryker Corp (SYK)} & \text{Consumer, Non-Cyclical} & \text{Health Care - Products} & \text{Medical Products}
\\ 434 & \text{VAR} & \text{Varian Medical Systems Inc (VAR)} & \text{Consumer, Non-Cyclical} & \text{Health Care - Products} & \text{Medical Products}
\\ 435 & \text{ZMH} & \text{Zimmer Holdings Inc (ZMH)} & \text{Consumer, Non-Cyclical} & \text{Health Care - Products} & \text{Medical Products}
\\ 436 & \text{HSIC} & \text{Henry Schein Inc (HSIC)} & \text{Consumer, Non-Cyclical} & \text{Health Care - Products} & \text{Medical Products}
\\ 437 & \text{DVA} & \text{DaVita HealthCare Partners Inc (DVA)} & \text{Consumer, Non-Cyclical} & \text{Health Care - Services} & \text{Dialysis Centers}
\\ 438 & \text{LH} & \text{Laboratory Corp of America Holdings (LH)} & \text{Consumer, Non-Cyclical} & \text{Health Care - Services} & \text{Medical Labs \& Testing Srv}
\\ 439 & \text{DGX} & \text{Quest Diagnostics Inc (DGX)} & \text{Consumer, Non-Cyclical} & \text{Health Care - Services} & \text{Medical Labs \& Testing Srv}
\\ 440 & \text{AET} & \text{Aetna Inc (AET)} & \text{Consumer, Non-Cyclical} & \text{Health Care - Services} & \text{Medical-HMO}
\\ 441 & \text{CI} & \text{Cigna Corp (CI)} & \text{Consumer, Non-Cyclical} & \text{Health Care - Services} & \text{Medical-HMO}
\\ 442 & \text{CVH} & \text{Coventry Health Care Inc (CVH)} & \text{Consumer, Non-Cyclical} & \text{Health Care - Services} & \text{Medical-HMO}
\\ 443 & \text{HUM} & \text{Humana Inc (HUM)} & \text{Consumer, Non-Cyclical} & \text{Health Care - Services} & \text{Medical-HMO}
\\ 444 & \text{UNH} & \text{UnitedHealth Group Inc (UNH)} & \text{Consumer, Non-Cyclical} & \text{Health Care - Services} & \text{Medical-HMO}
\\ 445 & \text{WLP} & \text{WellPoint Inc (WLP)} & \text{Consumer, Non-Cyclical} & \text{Health Care - Services} & \text{Medical-HMO}
\\ 446 & \text{THC} & \text{Tenet Healthcare Corp (THC)} & \text{Consumer, Non-Cyclical} & \text{Health Care - Services} & \text{Medical-Hospitals}
\\ 447 & \text{CLX} & \text{Clorox Co/The (CLX)} & \text{Consumer, Non-Cyclical} & \text{Household Products/Wares} & \text{Consumer Products-Misc}
\\ 448 & \text{KMB} & \text{Kimberly-Clark Corp (KMB)} & \text{Consumer, Non-Cyclical} & \text{Household Products/Wares} & \text{Consumer Products-Misc}
\\ 449 & \text{AVY} & \text{Avery Dennison Corp (AVY)} & \text{Consumer, Non-Cyclical} & \text{Household Products / Wares} & \text{Office Supplies \& Forms}
\\ 450 & \text{ABT} & \text{Abbott Laboratories (ABT)} & \text{Consumer, Non-Cyclical} & \text{Pharmaceuticals} & \text{Medical-Drugs}
\\ 451 & \text{AGN} & \text{Allergan Inc/United States (AGN)} & \text{Consumer, Non-Cyclical} & \text{Pharmaceuticals} & \text{Medical-Drugs}
\\ 452 & \text{BMY} & \text{Bristol-Myers Squibb Co (BMY)} & \text{Consumer, Non-Cyclical} & \text{Pharmaceuticals} & \text{Medical-Drugs}
\\ 453 & \text{FRX} & \text{Forest Laboratories Inc (FRX)} & \text{Consumer, Non-Cyclical} & \text{Pharmaceuticals} & \text{Medical-Drugs}
\\ 454 & \text{JNJ} & \text{Johnson \& Johnson (JNJ)} & \text{Consumer, Non-Cyclical} & \text{Pharmaceuticals} & \text{Medical-Drugs}
\\ 455 & \text{LLY} & \text{Eli Lilly \& Co (LLY)} & \text{Consumer, Non-Cyclical} & \text{Pharmaceuticals} & \text{Medical-Drugs}
\\ 456 & \text{MRK} & \text{Merck \& Co Inc (MRK)} & \text{Consumer, Non-Cyclical} & \text{Pharmaceuticals} & \text{Medical-Drugs}
\\ 457 & \text{PFE} & \text{Pfizer Inc (PFE)} & \text{Consumer, Non-Cyclical} & \text{Pharmaceuticals} & \text{Medical-Drugs}
\\ 458 & \text{MYL} & \text{Mylan Inc/PA (MYL)} & \text{Consumer, Non-Cyclical} & \text{Pharmaceuticals} & \text{Medical-Generic Drugs}
\\ 459 & \text{PRGO} & \text{Perrigo Co (PRGO)} & \text{Consumer, Non-Cyclical} & \text{Pharmaceuticals} & \text{Medical-Generic Drugs}
\\ 460 & \text{ACT} & \text{Actavis Inc (ACT)} & \text{Consumer, Non-Cyclical} & \text{Pharmaceuticals} & \text{Medical-Generic Drugs}
\\ 461 & \text{ABC} & \text{AmerisourceBergen Corp (ABC)} & \text{Consumer, Non-Cyclical} & \text{Pharmaceuticals} & \text{Medical-Whsle Drug Dist}
\\ 462 & \text{CAH} & \text{Cardinal Health Inc (CAH)} & \text{Consumer, Non-Cyclical} & \text{Pharmaceuticals} & \text{Medical-Whsle Drug Dist}
\\ 463 & \text{MCK} & \text{McKesson Corp (MCK)} & \text{Consumer, Non-Cyclical} & \text{Pharmaceuticals} & \text{Medical-Whsle Drug Dist}
\\ 464 & \text{ESRX} & \text{Express Scripts Holding Co (ESRX)} & \text{Consumer, Non-Cyclical} & \text{Pharmaceuticals} & \text{Pharmacy Services}
\end{array} \]

\normalsize

Table 5 - Tickers, company names, sectors, industries, and sub-industries of stocks used in this article.


\vskip 0.5 cm

\noindent {\bf \Large References}

\vskip 0.5 cm

\noindent Acemoglu, D., Osdaglar, A., and Tahbaz-Salehi, A. (2013). Systemic risk and stability in financial networks. The National Bureau of Economics Research Working Paper No. 18727.

\vskip 0.2 cm

\noindent  Allen, F. and Gale, D. (2000). Financial contagion. Journal of Political Economy 108, 1-33.

\vskip 0.2 cm

\noindent Allen, F. and Babus, A. (2009). Networks in finance. In Kleindorfer, P., Wing, Y.,and Gunther, R. (eds.), The network challenge: strategy, profit, and risk in an interlinked world. Wharton School Publishing.

\vskip 0.2 cm

\noindent Amini, H., Cont, R., Minca, A. (2012). Resilience to contagion in financial networks. ArXiv:1112.5687v1.

\vskip 0.2 cm

\noindent Ausloos, M. and Lambiotte, R. (2007). Clusters or networks of economies? A macroeconomy study through gross domestic product. Physica A 382, 16-21.

\vskip 0.2 cm

\noindent Baek, S.K., Jung, W-S, Kwon, O., Moon, H-T. (2005). Transfer Entropy Analysis of the Stock Market. ArXiv:physics/0509014v2.

\vskip 0.2 cm

\noindent Battiston, S., Puliga, M., Kaushik, R., Tasca, P., Caldarelli, G. (2012). DebtRank: too central to fail? Financial networks, the FED and systemic risk. Scientific reports 2, 1-6.

\vskip 0.2 cm

\noindent Battiston, S., Gatti, D.D., Gallegati, M., Greenwald, B., and Stiglitz. J. (2012a). Liaisons dangereuses: increasing connectivity, risk sharing, and systemic risk. Journal of Economic Dynamics and Control 36, 1121-1141.

\vskip 0.2 cm

\noindent Battiston, S., Gatti, D.D., Gallegati, M., Greenwald, B., and Stiglitz. J. (2012b). Default cascades: when does risk diversification increase stability? Journal of Financial Stability 8, 138-149.

\vskip 0.2 cm

\noindent Borg, I. and Groenen, P. (2005). Modern Multidimensional Scaling: theory and applications. 2nd edition, Springer-Verlag.

\vskip 0.2 cm

\noindent Boss, M., Elsinger, H., Summer, M., and Thurner, S. (2004). The network topology of the interbank market. Quantitative Finance 4, 677-684.

\vskip 0.2 cm

\noindent Buccheri, G., Marmi, S., and Mantegna, R. (2013). Evolution of correlation structure of industrial indices of US equity markets. Physical Review E 88, 01806.

\vskip 0.2 cm

\noindent Canedo, J.M.D., Mart\'\i nez-Jaramillo, S. (2010). Financial contagion: a network model for estimating the distribution of loss for the financial system. Journal of Economic Dynamics and Control 34, 2358-2374.

\vskip 0.2 cm

\noindent Castiglionesi, F., Navarro, N. (2007). Optimal fragile financial networks. Tilburg University Discussion Paper no. 2007-100.

\vskip 0.2 cm

\noindent Chinazzi, M., Fagiolo, G., Reyes, J.A., Schiavo, S. (2013). Post-mortem examination of the international financial network. Journal of Economic Dynamics and Control 37, 1692-1713.

\vskip 0.2 cm

\noindent Cossin, D., Schellhorn, H. (2007). Credit risk in a network economy. Management Science 53, 604-1,617.

\vskip 0.2 cm

\noindent Dimpfl, T. Huergo, L. and Peter, F.J. (2012). Using transfer entropy to measure information flows from and to the CDS market. Midwest Finance Association 2012 Annual Meetings Paper.

\vskip 0.2 cm

\noindent Dimpfl, T., Peter, F.J. (2012). Using transfer entropy to measure information flows between financial markets. Studies in Nonlinear Dynamics and Econometrics 17, 85-102.

\vskip 0.2 cm

\noindent Dimpfl, T, Peter, F.J. (2014). The impact of the financial crisis on transatlantic information flows: an intraday analysis. Journal of International Financial Markets, Institutions and Money 31, 1-13.

\vskip 0.2 cm

\noindent Elliott, M., Golub, B., Jackson, M.O. (2013). Financial Networks and Contagion. Available at SSRN: http://ssrn.com/abstract=2175056.

\vskip 0.2 cm

\noindent Gai, P., Haldane. A., Kapadia, S. (2011). Complexity, concentration and contagion. Journal of Monetary Economics 58, 453-470.

\vskip 0.2 cm

\noindent Gai, P., Kapadia, S. (2010). Contagion in Financial Networks. Proceedings of the Royal Society A 466.

\vskip 0.2 cm

\noindent Georg, C-P. (2010). The Effect of the Interbank Network Structure on Contagion and Financial Stability. Working Papers on Global Financial Markets, No. 12.

\vskip 0.2 cm

\noindent Haldane, A. (2009). Rethinking the financial network. Speech delivered at the Financial Student Association, Amsterdam, Aprol.

\vskip 0.2 cm

\noindent G. Hale. (2012). Bank relationships, business cycles, and financial crises. Journal of International Economics 88, 312-325.

\vskip 0.2 cm

\noindent Hattori, M and Suda, Y. (2008). Developments in a cross-border bank exposure ``network''. Bank of Japan Working Paper Series No.07-E-21.

\vskip 0.2 cm

\noindent Iori, G., Masi, G., Precup, O.V., Gabbi, G., Caldarelli, G. (2008). A network analysis of the Italian overnight money market. Journal of Economic Dynamics and Control 32, 259-278.

\vskip 0.2 cm

\noindent Jizba, P., Kleinert, H. and Shefaat, M. (2012). Renyi's information transfer between financial time series. Physica A 391, 2971-2989.

\vskip 0.2 cm

\noindent Kaushik, R. and Battiston, S. (2012) Credit Default Swaps drawup networks: too interconnected to be stable? PLOS ONE 8, e61815.

\vskip 0.2 cm

\noindent Kim, J., Kim, G, An, S., Kwon, Y-K., Yoon, S. (2013). Entropy-based analysis and bioinformatics-inspired integration of global economic information transfer. PLOS One 8, e51986.

\vskip 0.2 cm

\noindent Kirman, A (1997). The economy as an evolving network. Journal of Evolutionary Economics 7, 339-353.

\vskip 0.2 cm

\noindent Kubelec, C., S\'a, F. (2010). The geographical composition of national external balance sheets: 1980-2005. Bank of England Working Paper No.384.

\vskip 0.2 cm

\noindent Kwon, O. and Yang, J-S (2008a). Information flow between composite stock index and individual stocks. Physica A 387, 2851-2856.

\vskip 0.2 cm

\noindent Kwon, O. and Yang, J-S (2008b). Information flow between stock indices. European Physics Letters 82, 68003.

\vskip 0.2 cm

\noindent Laloux, L., Cizeau, P., Bouchaud, J-P, Potters, M. (1999). Noise dressing of financial correlation matrices, Physical Review Letters 83, 1467–1470.

\vskip 0.2 cm

\noindent Lee, K-M, Yang, J-S, Kim G., Lee J., Goh K-I, Kim, I-M. (2011). Impact of the Topology of Global Macroeconomic Network on the Spreading of Economic Crises. PLoS ONE 6, e18443.

\vskip 0.2 cm

\noindent Leitner, Y. (2005). Financial networks: contagion, commitment and private sector bailouts. Journal of Finance 60, 925-2,953.

\vskip 0.2 cm

\noindent Li, J., Liang, C., Zhu, X., Sun, X., Wu, D. (2013). Risk contagion in Chinese banking industry: a Transfer Entropy - based analysis. Entropy 15, 5549-5564.

\vskip 0.2 cm

\noindent Lorenz, J., Battiston, S., Schweitzer, F. (2009). Systemic risk in a unifying framework for cascading processes on networks. European Physics Journal B 71, 441-460.

\vskip 0.2 cm

\noindent Markose, S., Giansante, S., Gatkowski, M., Shaghaghi, A.R. (2010). Too Interconnected To Fail: Financial Contagion and Systemic Risk In Network Model of CDS and Other Credit Enhancement Obligations of US Banks. COMISEF Working Papers Series WPS-033 21/04/2010.

\vskip 0.2 cm

\noindent Mantegna, R.N. (1999). Hierarchical structure in financial markets. The European Physics Journal B 11, 193.

\vskip 0.2 cm

\noindent Marschinski, R., Kantz, H. (2002). Analysing the information flow between financial time series - an improved estimator for transfer entropy. The European Physical Journal B 30, 275-281.

\vskip 0.2 cm

\noindent Mart\'\i nez-Jaramillo, S., Alexandrova-Kabadjova, B., Bravo-Ben\'\i tez, B., Sol\'orzano-Margain, J.P. (2012). An empirical study of the Mexican banking system's network and its implications for systemic risk. Bank of Mexico Working Paper No. 2012-07.

\vskip 0.2 cm

\noindent Memmel, C. and Sachs, A. (2013). Contagion in the interbank market and its determinants. Journal of Financial Stability 9, 46-54.

\vskip 0.2 cm

\noindent Minoiu, C., Reyes, J.A. (2011). A network analysis of global banking: 1978-2009. IMF Working Paper WP/11/74, International Monetary Fund.

\vskip 0.2 cm

\noindent M\"uller, J. (2006). Interbank credit lines as a channel of contagion. Journal of Financial Services Research 29, 37-60.

\vskip 0.2 cm

\noindent Newman, M.E.J. (2010). Networks, and introduction. Oxford University Press.

\vskip 0.2 cm

\noindent Nier, E., Yang, J., Yorulmazer, T., Alentorn, A. (2007). Network models and financial stability. Journal of Economic Dynamics and Control 31,033-2060.

\vskip 0.2 cm

\noindent Onnela, J.-P., Chakraborti, A., Kaski, K. (2003). Dynamics of market correlations: taxonomy and portfolio analysis. Physical Review E 68, 1-12.

\vskip 0.2 cm

\noindent Onnela, J.-P., Chakraborti, A., Kaski, K., Kert\'{e}sz, J. (2002). Dynamic asset trees and portfolio analysis. The European Physics Journal B 30, 285-288.

\vskip 0.2 cm

\noindent Onnela, J.-P., Chakraborti, A., Kaski, K., Kert\'{e}sz, J. (2003). Dynamic asset trees and Black Monday. Physica A 324, 247-252.

\vskip 0.2 cm

\noindent Onnela, J.-P., Chakraborti, A., Kaski, K., Kert\'{e}sz, J. (2004). Clustering and information in correlation based financial networks. The European Physics Journal B 38, 353-362.

\vskip 0.2 cm

\noindent Onnela, J.-P., Chakraborti, A., Kaski, K., Kert\'{e}sz, J., Kanto, A. (2003). Asset trees and asset graphs in financial markets. Physica Scripta T 106, 48-54.

\vskip 0.2 cm

\noindent Peter, F.J., Dimpfl, T., Huergo, L. (2012) Using transfer entropy to measure information flows from and to the CDS market. In Midwest Finance Association 2012 Annual Meetings Paper.

\vskip 0.2 cm

\noindent Sandoval Jr., L. and Franca, I. De P. (2012). Correlation of financial markets in times of crisis. Physica A 391, 187–208.

\vskip 0.2 cm

\noindent Sandoval Jr., L. (2012). A Map of the Brazilian Stock Market. Advances in Complex Systems 15, 1250042-1250082.

\vskip 0.2 cm

\noindent Sandoval Jr., L. (2013). Cluster formation and evolution in networks of financial market indices. Algorithmic Finance 2, 3-43.

\vskip 0.2 cm

\noindent Sandoval Jr., L. (2014a). To lag or not to lag? How to compare indices of stock markets that operate at different times. Physica A 403, 227–243.

\vskip 0.2 cm

\noindent Sandoval Jr., L. (2014b). Structure of a Global Network of Financial Companies based on Transfer Entropy. Working paper, submitted to Entropy.

\vskip 0.2 cm

\noindent Sandoval Jr., L. and Kenett, D.Y. (2014). Causality relations among international stock market indices. Working paper.

\vskip 0.2 cm

\noindent Schreiber, T. (2000). Measuring information transfer. Physical Review Letters 85, 461-464.

\vskip 0.2 cm

\noindent Schweitzer, F., Fagiolo, G., Sornette, D., Vega-Redondo, F., White. D.R. (2009). Economic Networks: What do we know and what do we need to know? Advances in Complex Systems 12, 407-422.

\vskip 0.2 cm

\noindent Shannon, C.E. (1948). A Mathematical Theory of Communication. Bell System Technical Journal 27, 379–423, 623-656.

\vskip 0.2 cm

\noindent Sinha, S. and Pan R.K. (2007). Uncovering the internal structure of the Indian financial market: cross-correlation behavior in the NSE. In ``Econophysics of markets and business networks'', Springer, 215-226.

\vskip 0.2 cm

\noindent Soram\"aki, K., Bech, M.L., Arnold, J., Glass, R.J., Beyeler, W.E. (2007). The topology of interbank payment flows. Physica A 379, 317333.

\vskip 0.2 cm

\noindent Tabak, B.M., Takami, M., Rocha, J.M.C., Cajueiro, D.O. (2011). Directed Clustering Coefficient as a Measure of Systemic Risk in Complex Banking Networks. Working Papers Series, Central Bank of Brazil, Number 249.

\vskip 0.2 cm

\noindent Upper, C. (2011). Simulation methods to assess the danger of contagion in interbank markets. Journal of Financial Stability 7 (3), 111–125.

\vskip 0.2 cm

\noindent Vivier-Lirimont, S. (2004). Interbanking networks: towards a small financial world? Cahiers de la Maison des Sciences Economiques v04046. Université Panthon-Sorbonne (Paris 1).

\vskip 0.2 cm

\noindent Watts, D. (2002). A simple model of global cascades on random networks. Proceedings of the National Academy Sciences 99, 766-5,771.

\begin{figure}[H]
\begin{center}
\includegraphics[scale=0.8]{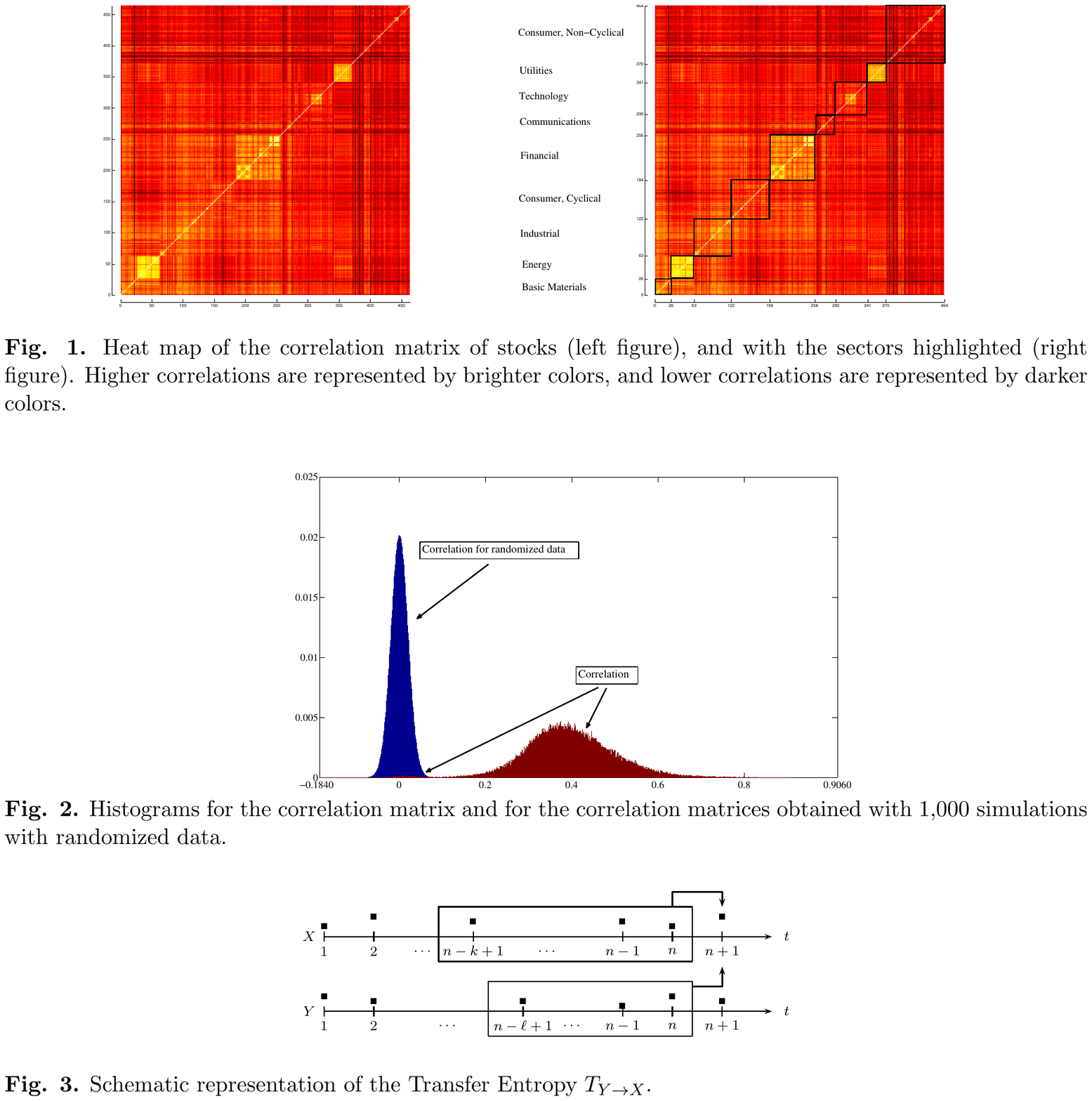}
\end{center}
\end{figure}

\begin{figure}[H]
\begin{center}
\includegraphics[scale=0.8]{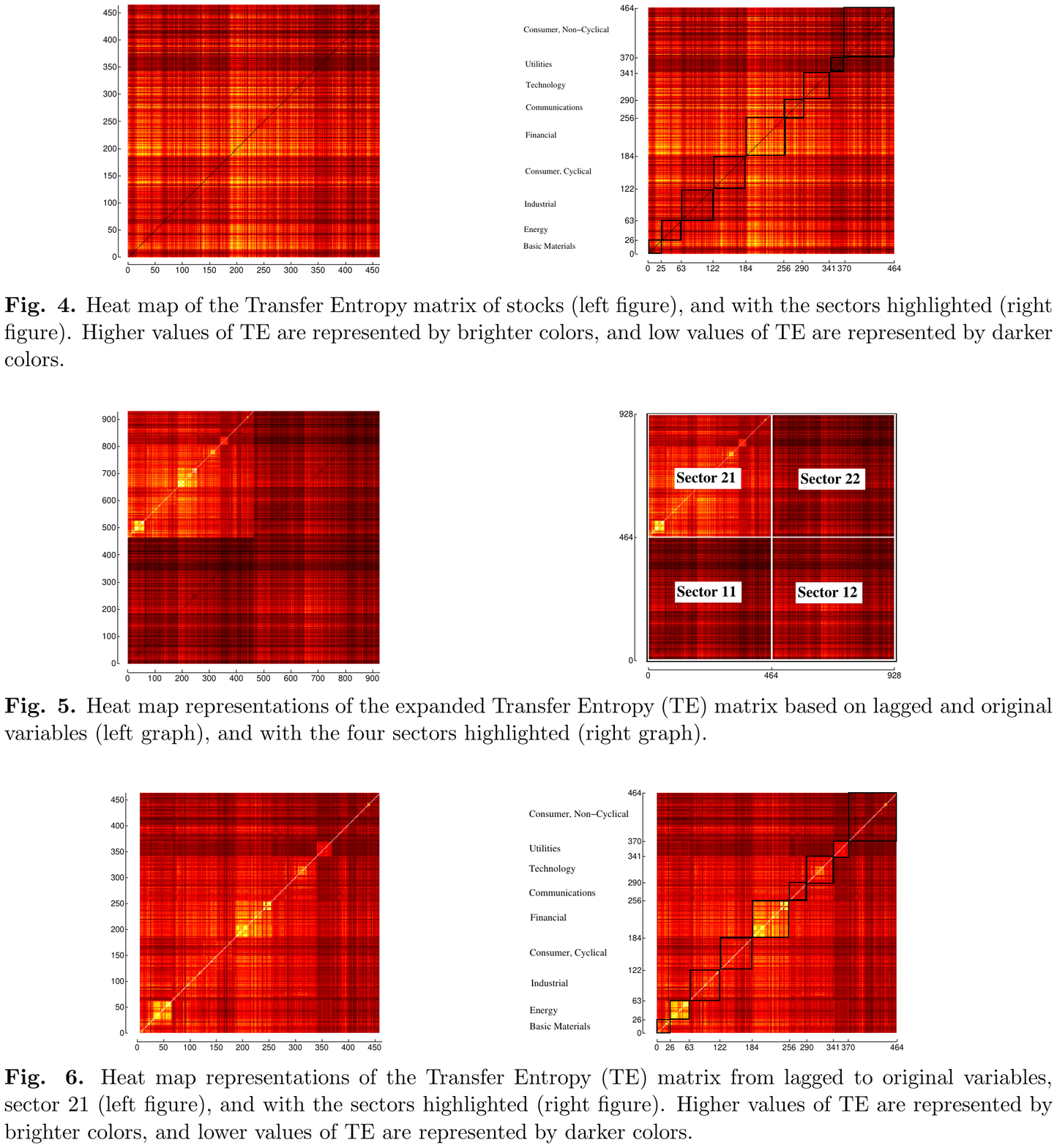}
\end{center}
\end{figure}

\begin{figure}[H]
\begin{center}
\includegraphics[scale=0.8]{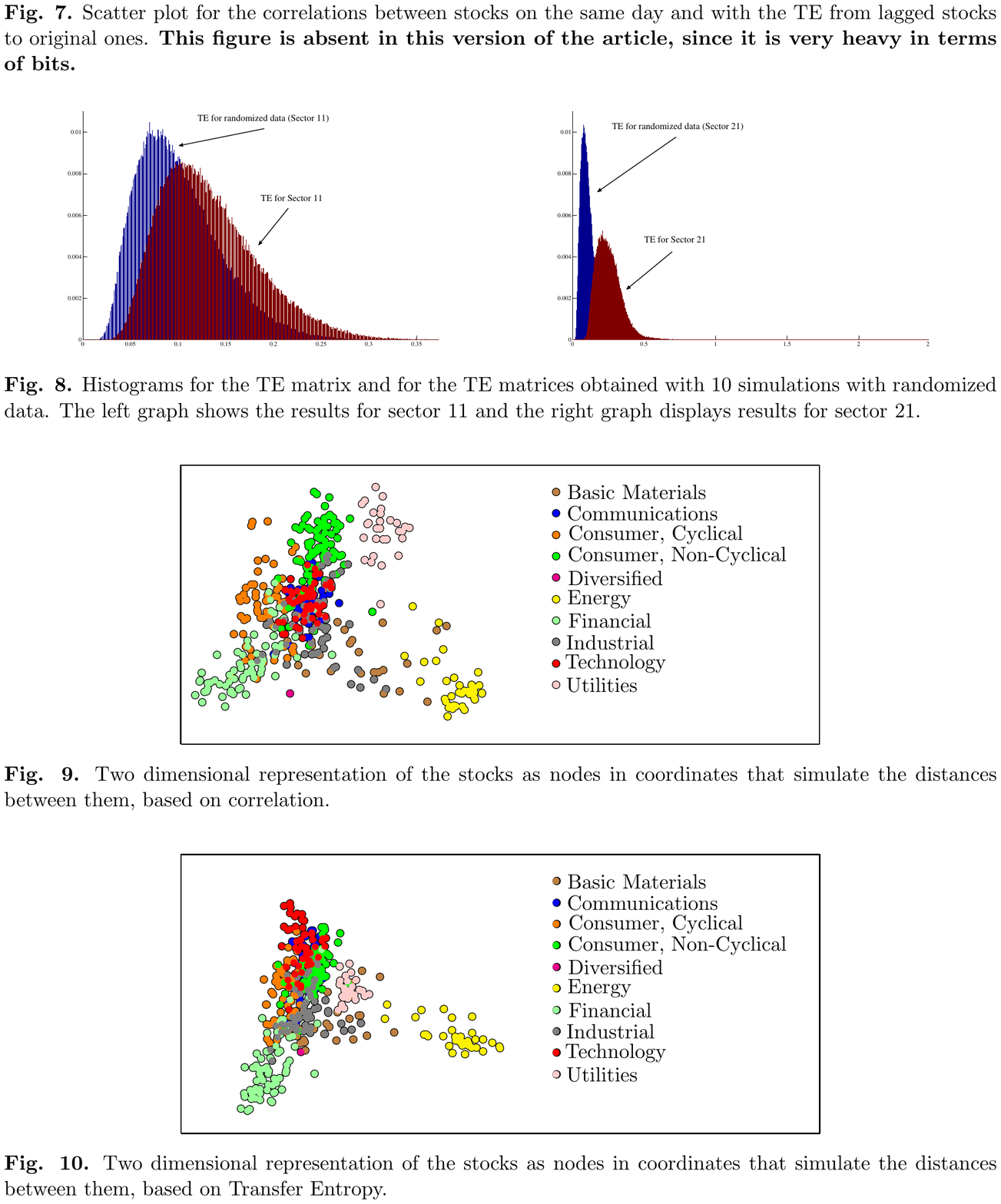}
\end{center}
\end{figure}

\begin{figure}[H]
\begin{center}
\includegraphics[scale=0.8]{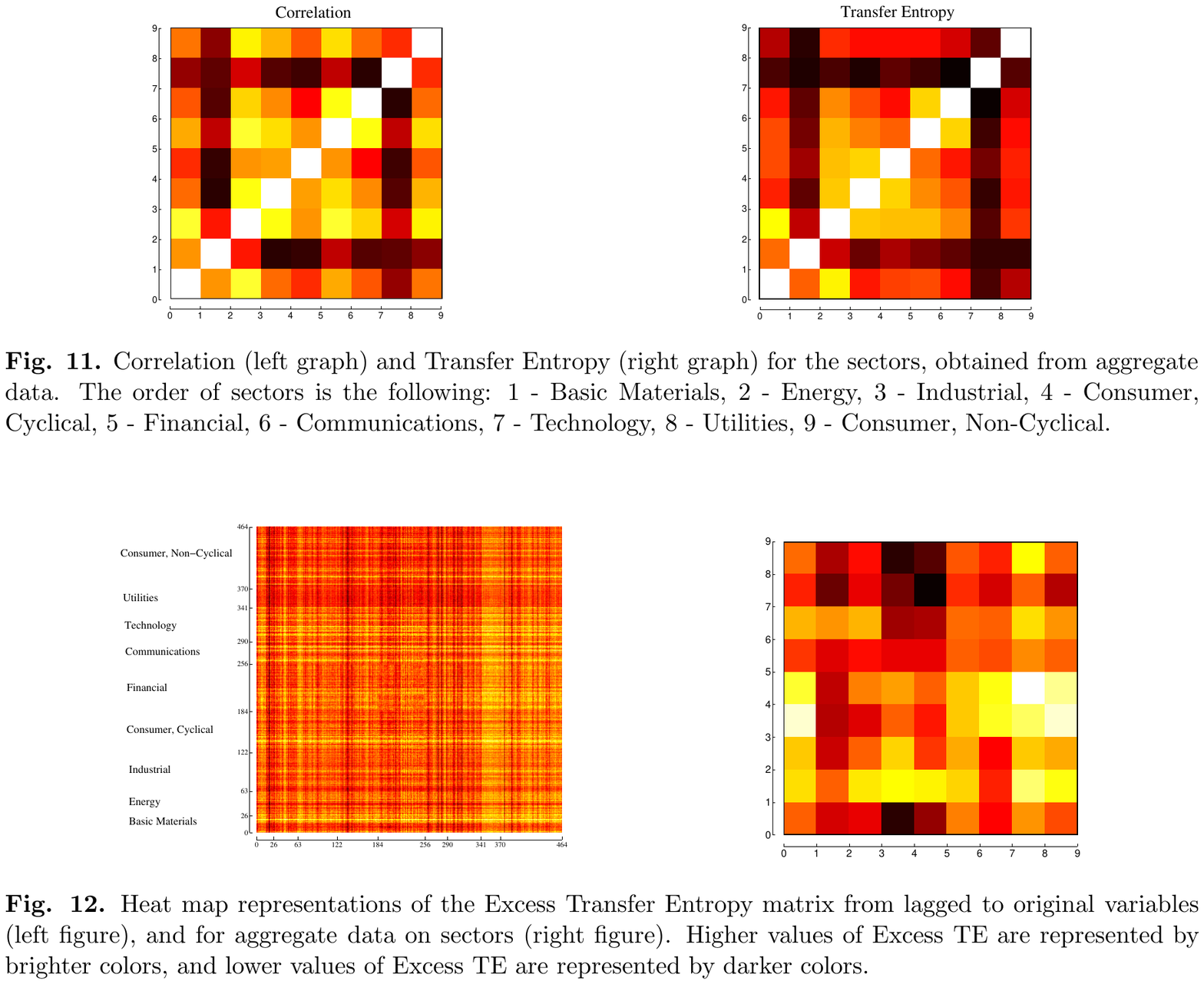}
\end{center}
\end{figure}

\begin{figure}[H]
\begin{center}
\includegraphics[scale=0.8]{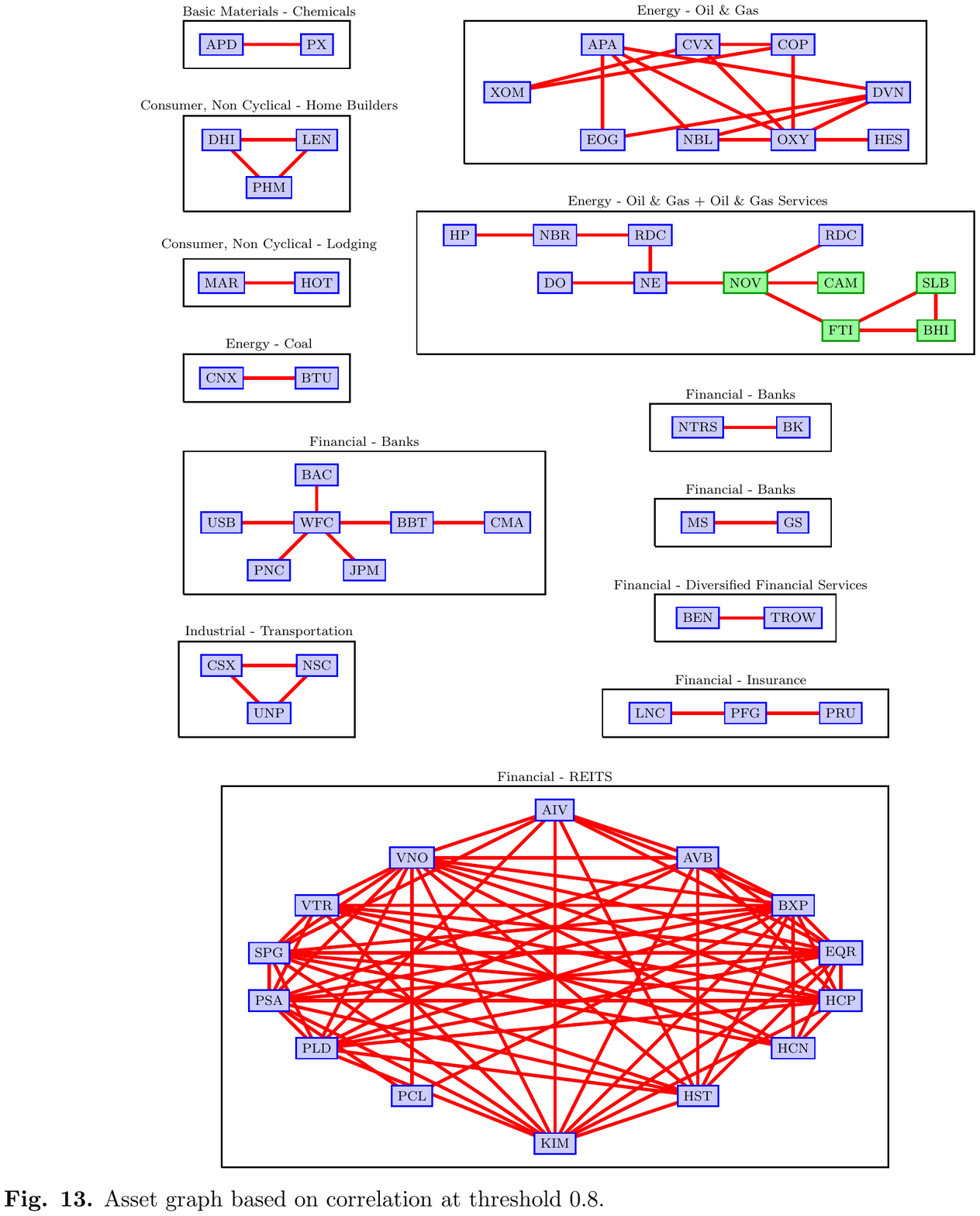}
\end{center}
\end{figure}

\begin{figure}[H]
\begin{center}
\includegraphics[scale=0.8]{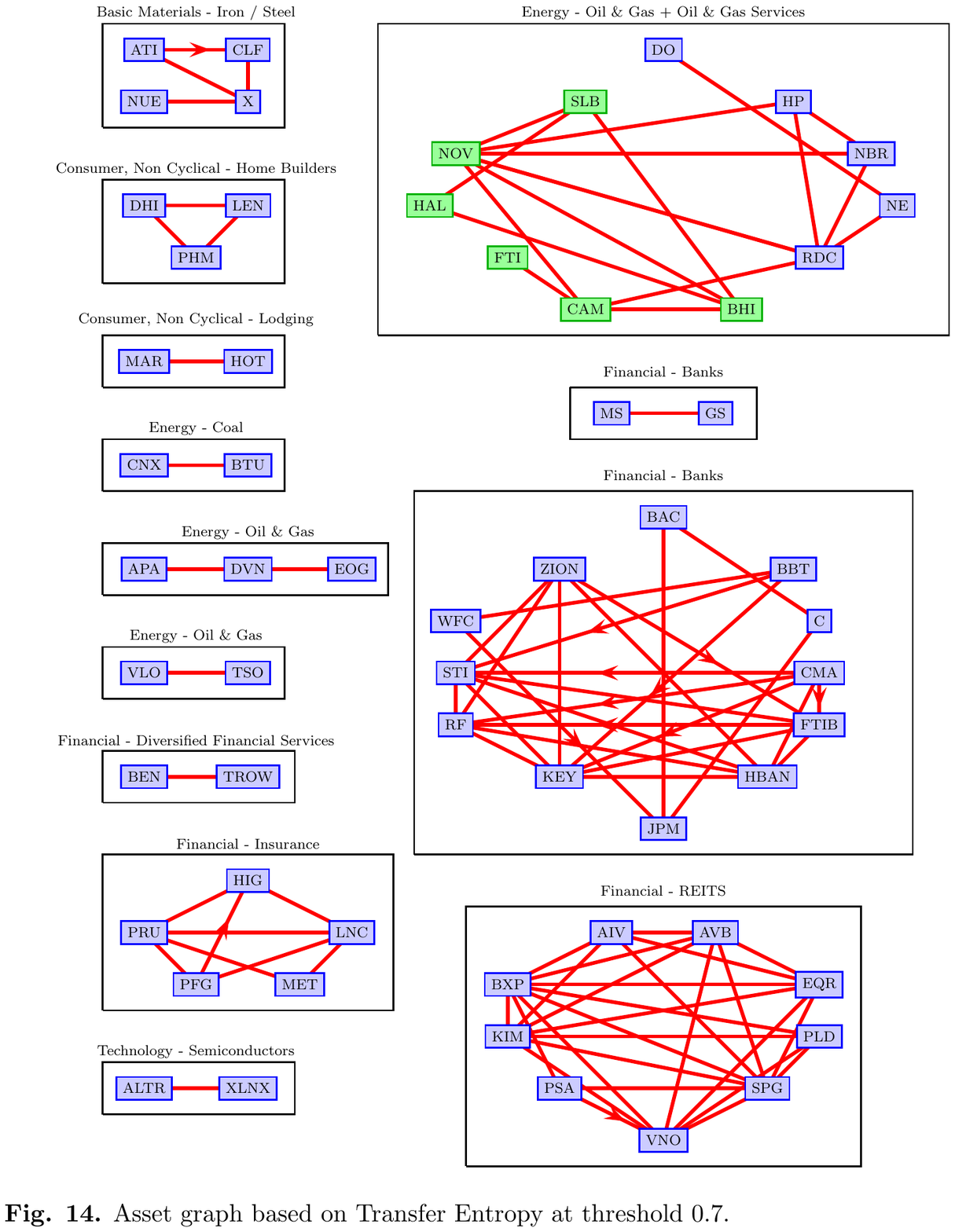}
\end{center}
\end{figure}

\begin{figure}[H]
\begin{center}
\includegraphics[scale=0.8]{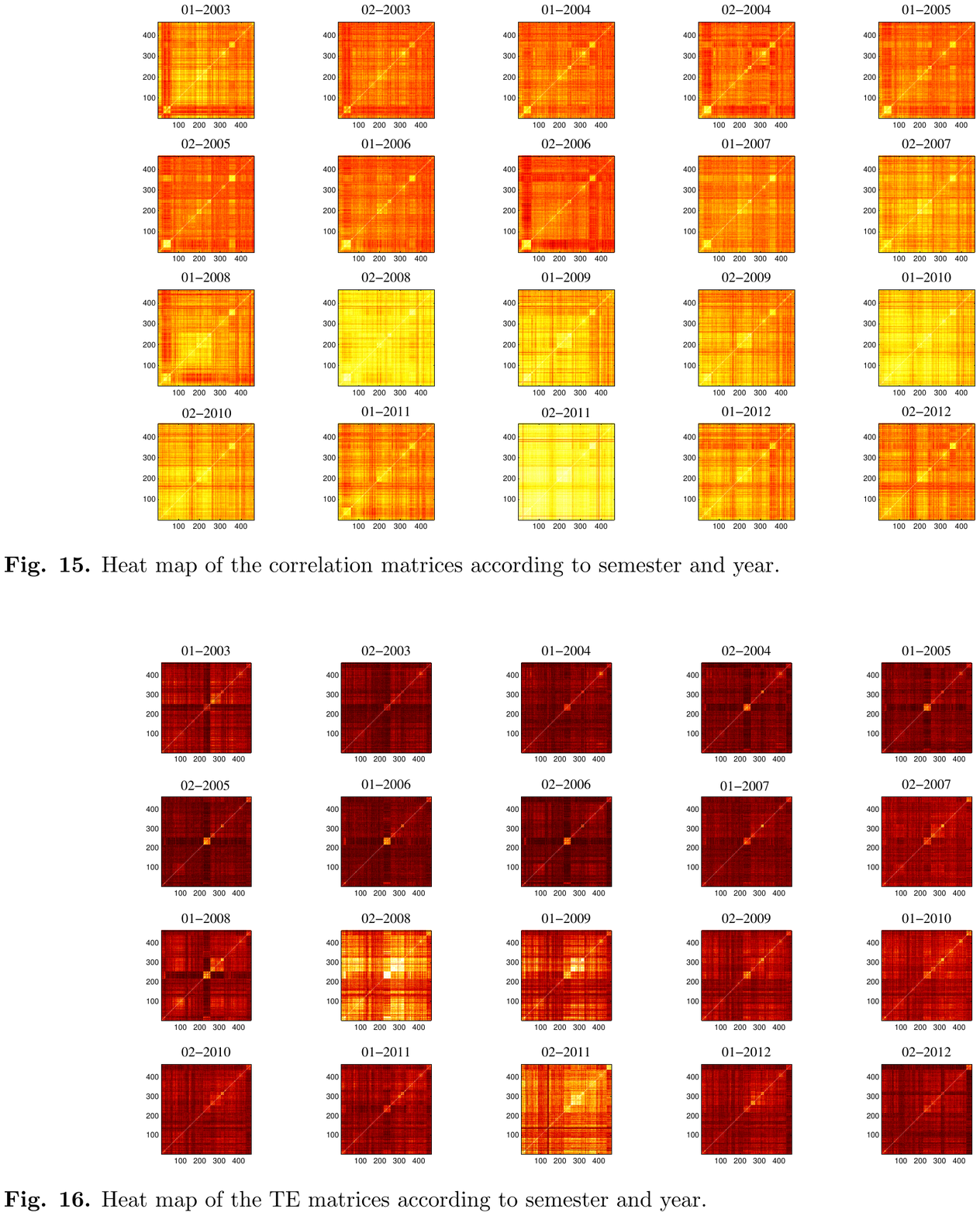}
\end{center}
\end{figure}

\begin{figure}[H]
\begin{center}
\includegraphics[scale=0.8]{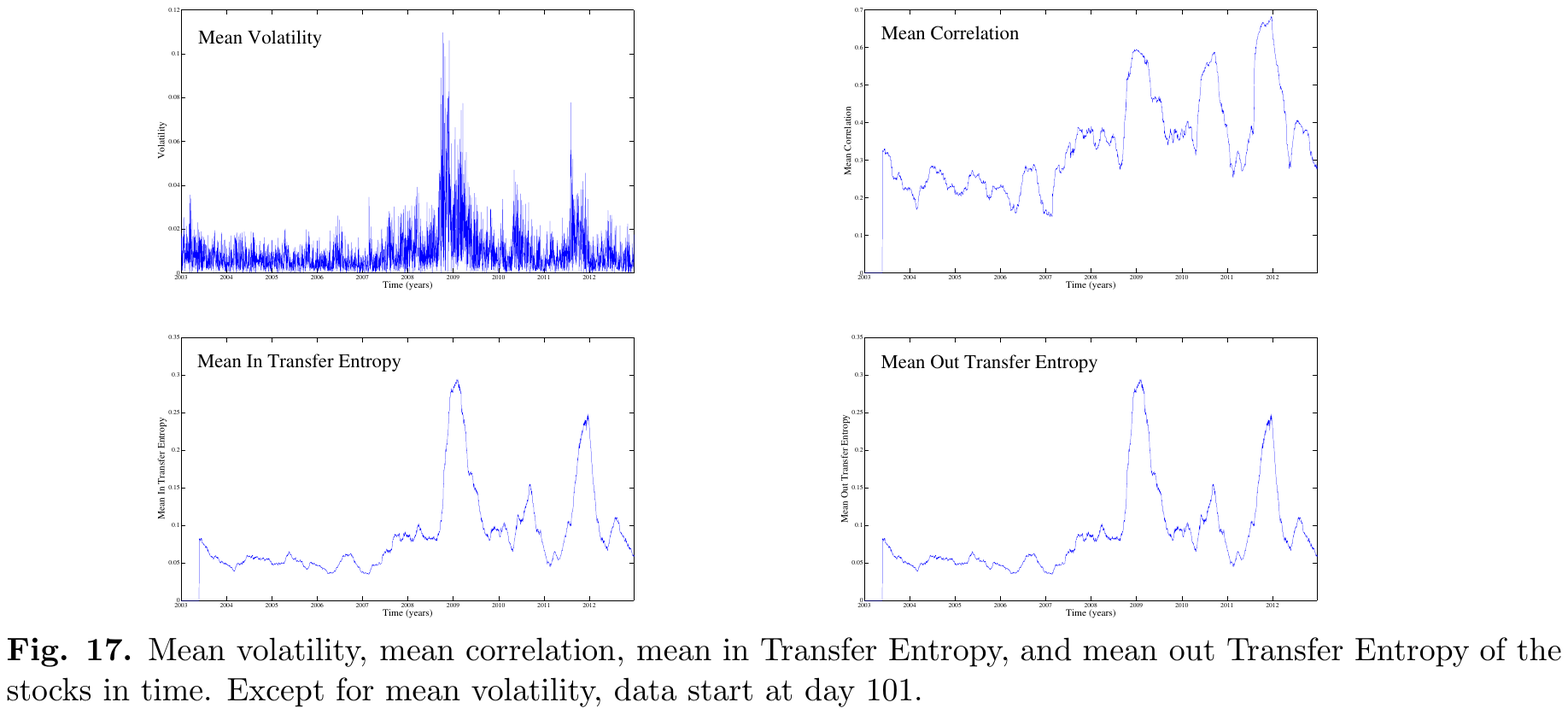}
\end{center}
\end{figure}

\begin{figure}[H]
\begin{center}
\includegraphics[scale=0.8]{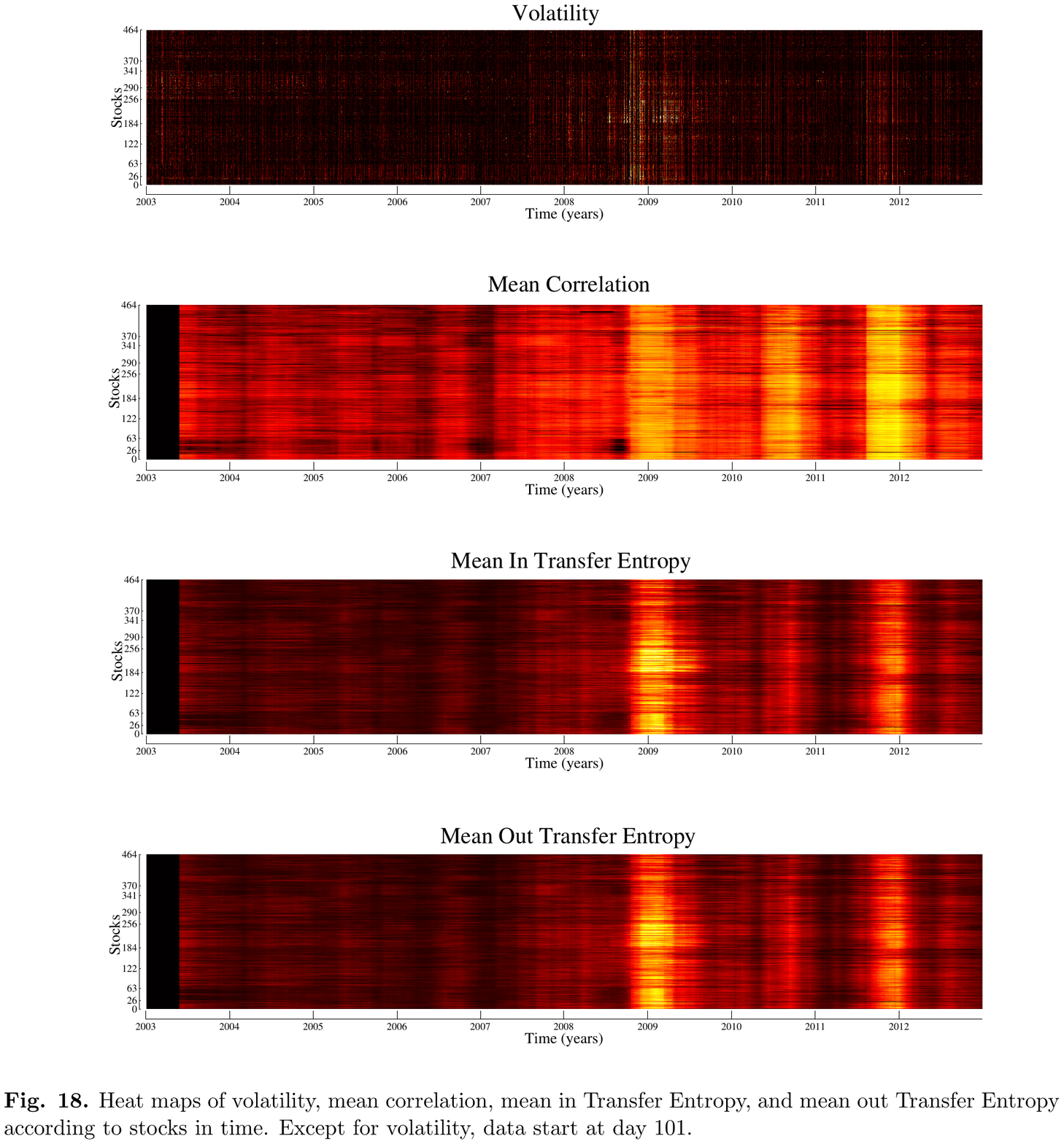}
\end{center}
\end{figure}

\begin{figure}[H]
\begin{center}
\includegraphics[scale=0.8]{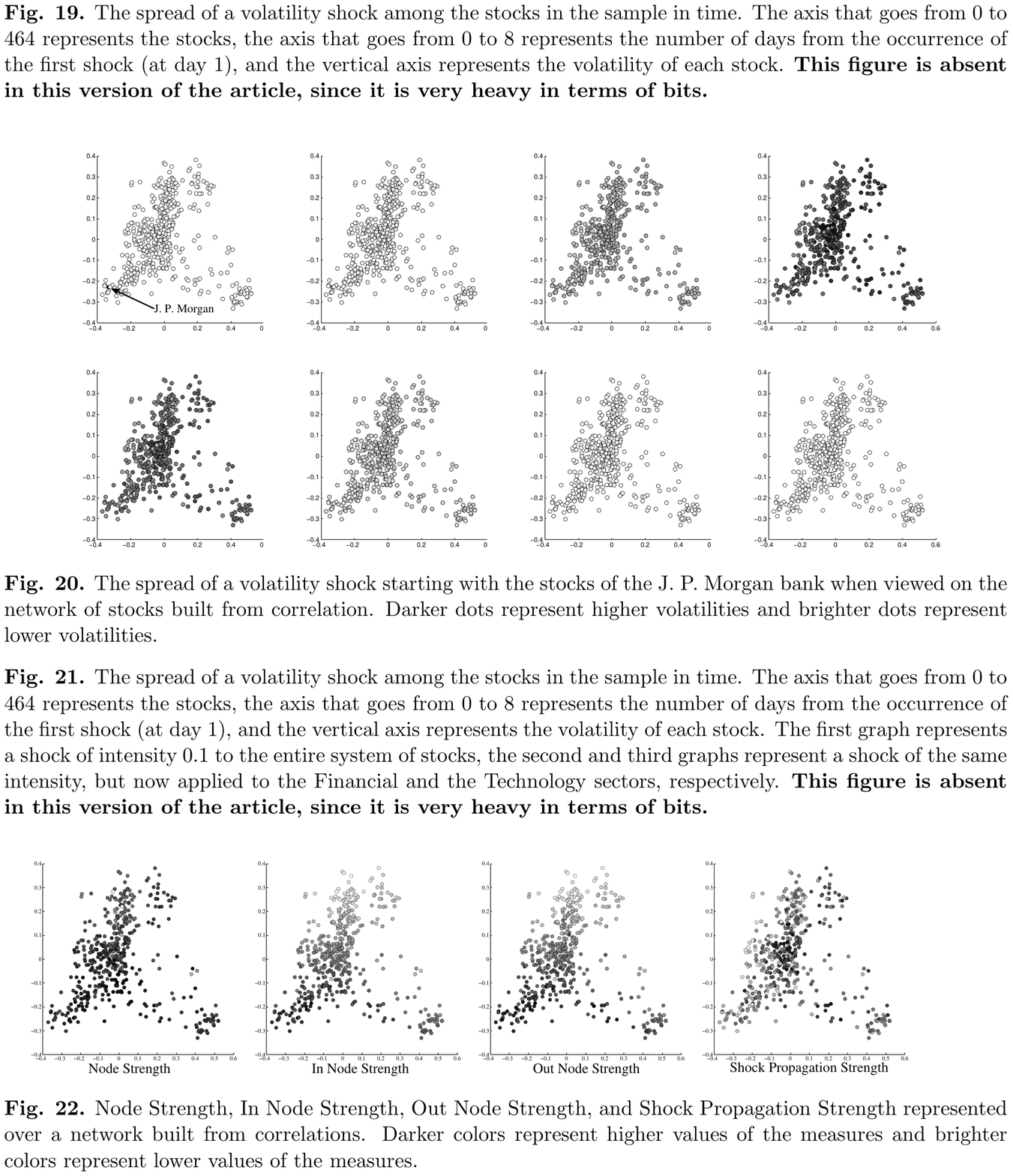}
\end{center}
\end{figure}

\end{document}